\documentclass[article, moreauthors,pdftex, nolinenumber]{Definitions/mdpi1} 
\preto{\abstractkeywords}{\nolinenumbers}
\usepackage{siunitx}
\usepackage{color}
\usepackage{amssymb}
\usepackage[normalem]{ulem}

\global\long\def\mr#1{\mathrm{#1}}
\global\long\def\qO{q_\mathrm{O}}

\global\long\def\kt{k_B T}
\global\long\def\Va{V_\text{ext}}
\global\long\def\a{a_p} 

\firstpage{1} 

\Title{Diffusion Limitations and Translocation Barriers in Atomically Thin Biomimetic Pores}

\Author{{Subin Sahu} 
 $^{1,2,3}$\orcidA{} and Michael Zwolak $^{1,}$*\orcidB{} }
\AuthorNames{Subin Sahu and Michael Zwolak}
\address{$^{1}$ \quad Biophysical and Biomedical Measurement Group, Microsystems and Nanotechnology Division, Physical Measurement Laboratory, National Institute of Standards and Technology, Gaithersburg, MD 20899, USA \\ 
 $^{2}$ \quad Institute for Research in Electronics and Applied Physics and Maryland NanoCenter, University of Maryland, College Park, MD 20742, USA\\
 $^{3}$ \quad Department of Chemical and Biological Engineering, University of Colorado Boulder, Boulder, CO 80309, USA}
\corres{Correspondence: mpz@nist.gov}

\abstract{
Ionic transport in nano- to sub-nano-scale pores is highly dependent on translocation barriers and potential wells. These features in the free-energy landscape are primarily the result of ion dehydration and electrostatic interactions. For pores in atomically thin membranes, such as graphene, other factors come into play. Ion dynamics both inside and outside the geometric volume of the pore can be critical in determining the transport properties of the channel due to several commensurate length scales, such as the effective membrane thickness, radii of the first and the second hydration layers, pore radius, and Debye length. In particular, for biomimetic pores, such as the graphene crown ether we examine here, there are regimes where transport is highly sensitive to the pore size due to the interplay of dehydration and interaction with pore charge. Picometer changes in the size, e.g., due to a minute strain, can lead to a large change in conductance. Outside of these regimes, the small pore size itself gives a large resistance, even when electrostatic factors and dehydration compensate each other to give a relatively flat---e.g., near barrierless---free energy landscape. The permeability, though, can still be large and ions will translocate rapidly after they arrive within the capture radius of the pore. This, in turn, leads to diffusion and drift effects dominating the conductance. The current thus plateaus and becomes effectively independent of pore-free energy characteristics. Measurement of this effect will give an estimate of the magnitude of kinetically limiting features, and experimentally constrain the local electromechanical conditions. 
}


\begin{document}

\section{Introduction}
Ionic transport through nano- and sub-nano-scale pores elicits a tremendous amount of interest due to its relevance in cellular processes including neurotransmission, muscle contraction, and other biological processes~\cite{hille2001,zheng2015handbook}, as well as its application in technologies such as  desalination~\cite{abraham2017, Surwade2015, tanugi2012}, osmotic power generation~\cite{feng2016single,walker2017, graf2019light}, and bio-chemical sensing~\cite{zwolak_colloquium:_2008,heerema2016, deamer2016three, si2017nanopore, danda2019two}. Ref.~\cite{sahu2019colloquium} provides a recent review of the use of pores in 2D materials for these applications. Understanding transport mechanisms, particularly in biological settings, has remained challenging due to their complexity and dependence on atomic details~\cite{faucher2019critical}. Furthermore, even for uncharged membranes, the region outside the pore can play a significant role in determining ionic transport, either via access resistance~\cite{sahu2018maxwell, sahu2018golden,aguilella2020access} or via diffusion limitations~\cite{hille1970ionic}. 

Synthetic nanopores offer the ability to study the factors that underpin transport mechanisms, such as the role of dehydration~\cite{zwolak2009quantized,song2009, zwolak2010dehydration, sahu2017dehydration, sahu2017ionic} and functional groups~\cite{sint2008,he2013bioinspired, sahu2019optimal,guardiani2019exploring}, and give rise to functional behavior, such as ion selectivity~\cite{sahu2019optimal,gibby2018role}. Pores in 2D membranes, in particular, have a larger access resistance, compared to their pore resistance, due to their small aspect ratio, $\a/h_p$, where $\a$ is the effective pore radius and $h_p$ is the effective pore length. Therefore, 2D materials provide a unique opportunity to study geometric effects in transport~\cite{liu2017geometrical}, atomic changes in area via precise control in pore fabrication and height via layering~\cite{sahu2017ionic}, and the interplay of various length scales relevant to the problem~\cite{sahu2019colloquium}, which will help in the design of separation membranes~\cite{sahu2019colloquium, bocquet2020nanofluidics, epsztein2020towards} and delineating factors relevant to biological channels~\cite{sahu2019colloquium}. In particular, the effective thickness is not the geometric thickness of the membrane (e.g., for graphene, twice the van der Waals radius of the carbon atoms), since ion size, hydration layer radii, and even membrane charge and the build up of charge layers give rise to an effective thickness~\cite{sahu2018maxwell}. When both the bulk and pore are within the continuum drift-diffusion regime, measuring or calculating the dependence of the conductance on pore radius quantifies the effective pore length through the equation
\begin{equation}\label{eq:TotalRes}
R=\gamma \left( \frac{1}{2\a} + \frac{h_p}{\pi \, \a^2} \right),
\end{equation}
which assumes a homogeneous resistivity $\gamma$ and a cylindrical pore geometry. The former entails that there are no concentration gradients and that the medium in the pore has the same resistivity as the bulk (otherwise, it requires an independent determination of the pore resistivity). For graphene, this approach yields an effective pore thickness of about 1~nm, both experimentally~\cite{garaj2010} (for unknown rim functionalization) and  computationally~\cite{sahu2018maxwell,sahu2019colloquium} (for unfunctionalized rims), provided that simulations properly include the influence of the bulk via the golden aspect ratio or associated scaling analysis~\cite{sahu2017golden,sahu2018maxwell}, as well as properly determine the pore radius~\cite{sahu2018maxwell}. The effective pore length is mostly due to the van der Waals radii and first hydration layer of the ions, which are both reflected in a build up of charge layers about 0.5~nm from the membrane.

More recently, it was demonstrated that applying strain to 2D pores can elucidate the conditions under which optimal transport and ion selectivity arise by modulating the balance of dehydration and electrostatic interactions~\cite{sahu2019optimal}. It  is unclear, however, whether 2D pores, and synthetic pores more generally, offer a means to investigate diffusion limitations. These arise due to fine details of pore structure typically thought to be out of our control. We will demonstrate here that the control provided by strain can tune atomically thin biomimetic pores into a diffusion-limited regime. Finding transitions into these regimes will help delineate and probe the electromechanical environment of nanopores, and elucidate diffusion-limited phenomena in more complex, biological settings. 

Ionic transport through a pore becomes diffusion-limited when the permeability of ions in the pore is large and the current is only restricted by the rate of diffusion of the ions from the bulk to the pore mouth~\cite{lauger1976, gates1990analytical}. In this diffusion-limited regime, the current does not increase with voltage as expected from Ohm's law. This is similar to the diffusion-limited processes in chemical reactions~\cite{collins1949diffusion} and other transport processes~\cite{behr1985carrier,wanunu2010electrostatic}. Diffusion-limited ionic currents are a regular occurrence in biological pores~\cite{brelidze2005probing, schroeder2007saturation}, as these can have the necessary conditions: narrow channels with high permeability for specific ions~\cite{hille2001} and the presence of ``inert'' ions~\cite{lauger1973ion}. Pores in atomically thin membranes, such as graphene, MoS$_2$, and hBN, also provide a very high permeability for ions due to their sub-nanoscale channel length~\cite{sahu2019colloquium}. Thus, in these membranes, the diffusion of the ions from the bulk to the mouth of the pore may become the limiting factor in the ion transport. It is the objective of this work to determine under what conditions bulk diffusion becomes the limiting factor, particularly when drift is also present. 

For the transport to be diffusion-limited, however, the drift contribution to the current in the bulk has to be small. This condition is hard to realize in pores in 2D materials under electrically driven transport, since a large portion of the applied voltage drops in the bulk solution, which in general has a higher resistance---in the form of access resistance---than the pore itself~\cite{garaj2010,sahu2018maxwell, sahu2018golden,sahu2019colloquium}. This is also true of small aspect ratio---short and wide---biological channels, where access resistance becomes dominant at low ion concentration~\cite{aguilella2020access}. There are regimes, though, where diffusion limitations may appear. Sub-nanoscale pores in graphene, for instance, have a high pore resistance~\cite{sahu2017dehydration}. As a result, most of the applied potential will drop across the pore, thus diminishing the drift current in the bulk and giving an opportunity to observe the diffusion-limited transport. At this sub-nanometer length scale, the translocation barriers and potential wells due to ion dehydration and electrostatic interactions play a major role in determining transport through such pores~\cite{sahu2019optimal}. When one or the other interaction dominates, translocation through the pore is barrier-limited. 

Under the right conditions---determined by pore size and charge, dehydration energy, etc.---the permeability of the pore will be large~\cite{berezhkovskii2005optimizing, kasianowicz2006enhancing} (e.g., ion channels with binding sites, in particular, can have an inverse relationship between permeability and conductance~\cite{bormann1987mechanism}). To understand the conditions for having a small pore conductance and high pore permeability, one can look at the continuum-limit~expressions
\begin{equation}
G_p= q\, c_p\, \mu_p\, A_p/h_p 
\end{equation}
and
\begin{equation}
P_p=D_p/h_p ,
\end{equation}
where $c_p$ is the ion concentration in the pore, $q$ is the charge of the ion, $\mu_p$ is the pore mobility, $D_p$ ($=\mu_p k_B T/q$) is the diffusion coefficient, $A_p$ is the pore area, $h_p$ is the pore length, $k_B$ is Boltzmann's constant, and $T$ is the temperature. We briefly note that all pore quantities (which have a subscript $p$) are effective quantities, as will be abundantly clear throughout this work. An overall barrier in the pore can limit the concentration by exponentially reducing the partition coefficient into the pore. When such a barrier is present, without features internal to the pore, the mobility and diffusion can be unaffected. Similarly, the internal features, and hence mobility, can be altered with very small relative changes to pore size (and vice versa, reducing the cross-sectional area of the pore can reduce the conductance but with little effect on mobility when the pore size is relatively large). Essentially, strain and voltage will give the right knobs to tune some pores into a diffusion-limited regime by modulation of $c_p$ and $\mu_p$, while retaining a small $A_p$ (i.e., a large pore resistance).  

\begin{figure}[h]
\centering
 \includegraphics[scale=1]{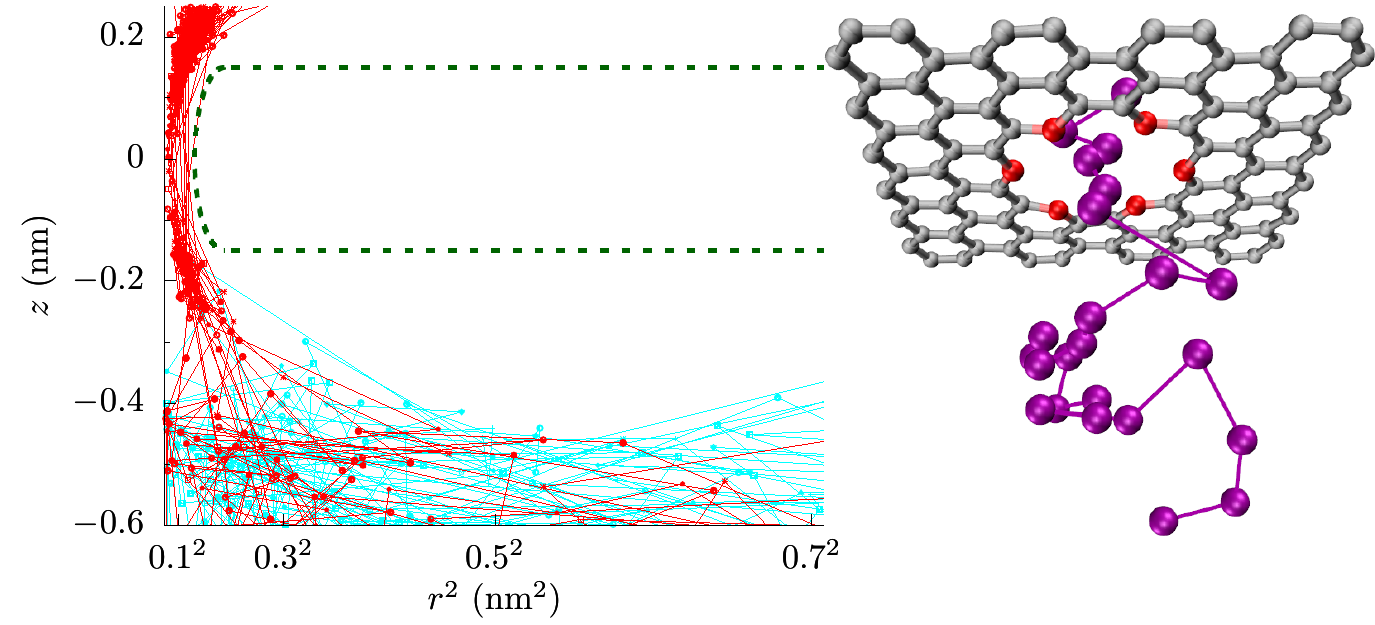}
 \caption{{\textbf{Ion trajectories in a graphene crown ether pore.}} 
 The trajectories of K$^+$ ions around the pore with $\qO=-0.54\,e$ at 4\% strain, with an applied bias of 0.25 V along the $z$-axis (making K$^+$ move in the positive $z$-direction). We plot the trajectories in $z$-$r^2$ space to keep the representation of volume constant. The red trajectories are for ions that translocate through the pore and the cyan trajectories are for ions that reflect back. The green dashed line shows the geometric boundary of the membrane determined by the van der Waals radii of the pore atoms. The effective boundary extends to $|z| \lesssim $~0.5~nm due to the size of the hydrated ions. The inset shows a portion of the graphene crown ether (red and grey spheres are oxygen and carbon atoms,  respectively) and a three-dimensional trajectory of a potassium ion (purple spheres, separated in time by 10~ps, connected with purple lines) crossing the pore. Connecting lines are a guide to the eye only. \label{fig:K-traj}}
\end{figure}

Here, we will examine the \mbox{18-crown-6} pore in graphene (see Figure~\ref{fig:K-traj}) under the influence of a homogeneous strain in the plane of the membrane and cross-membrane voltage for different local electronic pore environments. Crown ether pores were seen by scanning transmission electron microscopy of graphene membranes that were made by exfoliation of graphite~\cite{guo2014crown}. Even though strain only changes the pore area by a minuscule amount, the change in the free-energy barrier and hence the ionic current is substantial~\cite{sahu2019optimal}. We find both barrier-limited and diffusion-limited regimes depending on the strain, voltage, and local environment. In the barrier-limited regime, the conductance increases with applied voltage, as it helps ions overcome the barrier. In the diffusion-limited regime, the pore conductance decreases with voltage as the bias depletes the charge carriers in the pore and at its entrance. The transition between these regimes depends on the charge separation in the pore (e.g., the local dipole moment at pore rim), which is not experimentally known, nor is there sufficient thermodynamic or kinetic data from experiment to constrain it. Measurement of the pore conductance versus strain and voltage, therefore, gives a possible route to determining the electromechanical environment, and thus constraining the magnitude of charge separation at the pore rim. 

\section{Methods}
We performed all-atom molecular dynamics (MD) simulations using the NAMD2 simulations package~\cite{phillips2005}. The details of the simulations were the same as in Ref.~\cite{sahu2019optimal}. We applied a voltage between 0.1~V to 1.0~V and calculated the ionic current for a 1 mol/L KCl solution by counting the ions that crossed the pore. Since the pore rim is negatively charged and sub-nanoscale in size (i.e., the electrostatic interactions are not significantly screened), only cation currents were present in all cases. We calculated the free-energy barrier using the adaptive biasing force (ABF) method~\cite{Darve2008} in a cylinder of radius 0.28~nm and height 3~nm, centered at the pore. A portion of our simulation cell and a set of ion trajectories are shown in Figure~\ref{fig:K-traj}. We employed the golden aspect ratio method, as it is the only method that can properly capture bulk access effects~\cite{sahu2018maxwell,sahu2018golden}, and without it, one cannot explore bulk diffusion limitations with all-atom~simulations.

We calculated the ionic current and free-energy barrier at various homogeneous strains from 0\% to 10\% on the graphene membrane (we note that most of the features we observe occur at strains of 4\% to 6\%, and graphene can survive strains above 20\%~\cite{lee2008measurement,pereira2009tight}). The strain was within the membrane and thus tended to enlarge the pore (albeit by small amounts) and expand in-plane distances between atoms. In the unstrained pore, the nominal pore radius (measured from the pore center to the center of the oxygen atoms) was approximately 0.29~nm and increased by about 7.5~pm for each 1\% strain (this reflects a small---a factor of $\approx$2---geometric amplification~\cite{sahu2019optimal}). This yielded nominal pore sizes from 0.29~nm (at 0\% strain) to about 0.37~nm (at 10\% strain), with a roughly linear relationship with strain. Though the change in pore size was minuscule, the energy landscape changes substantially. The landscape also depended on the pore charge, which is, however, not known. For the crown ether pore in graphene, the charge per oxygen atom ($\qO$) could be between $-0.2\,e$ and $-0.7\,e$~\cite{sahu2019optimal, heath2018first, glendening1994ab, fang2019highly}. We thus used two representative test charges, $-0.24\,e$ and $-0.54\,e$. In the former, there was an energy barrier, and in the latter, there was a potential well at the center of the unstrained pore. Each of the 12 carbon atoms---the ones adjacent to the six oxygen atoms in the pore---had charge $-\qO/2$, and the rest of the carbon atoms were neutral. When we refer to the pore charge, we are referencing the local polarization of charge from the carbon atoms of the graphene near the pore and the oxygen atoms on the pore rim. For each data point (i.e., for a particular value of $\qO$, strain, and voltage), we performed five parallel production runs for a total simulation between 250~ns to 500~ns. This allowed for an error estimation using the standard error, $\text{SE}=\sqrt{\text{var}/n_r}$, where var is the variance between the $n_r=5$ parallel runs.

\section{Results}

\begin{figure}[H]
\centering
\includegraphics{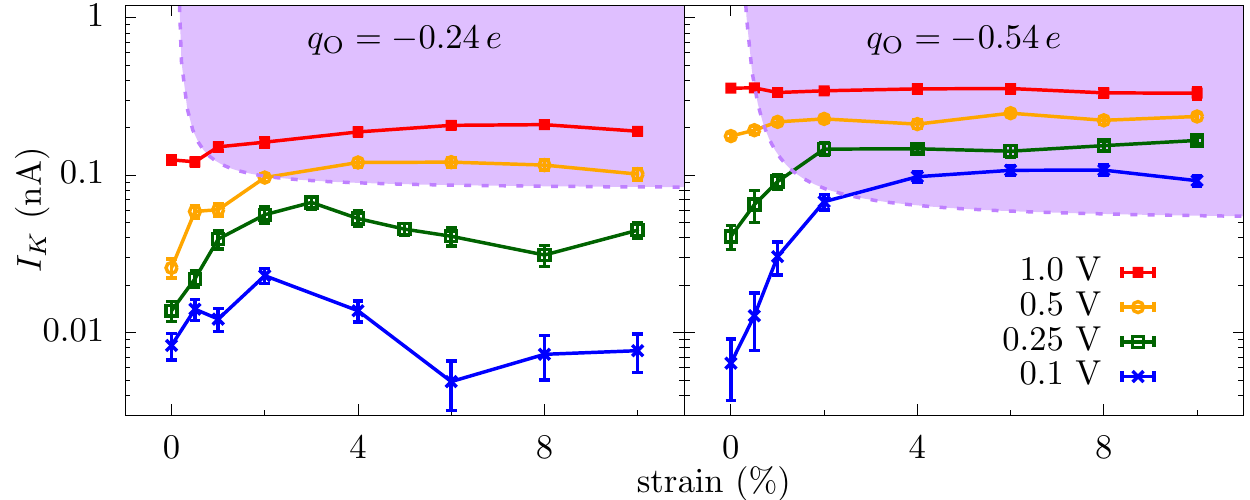}
\includegraphics{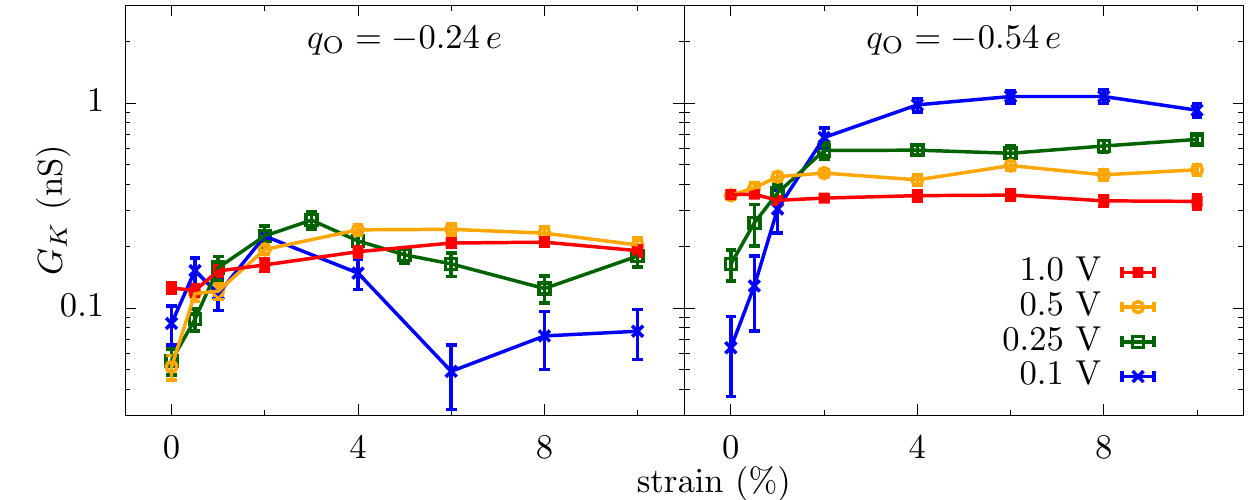}
\caption{{{
\bf Colossal rise, non-monotonicity, and saturation of the ionic current.}} 
 (\textbf{top panel}) Potassium current ($I_\mr{K}$) versus strain at various voltages across the graphene crown ether pore within 1 mol/L KCl. At small voltages and minute strains,  $I_\mr{K}$ increases rapidly with strain due to the large electromechanical susceptibility of the pore~\cite{sahu2019optimal}. A further increase of the strain causes $I_\mr{K}$ to either decrease (for $\qO=-0.24\,e$) or saturate (for $\qO=-0.54\,e$), albeit the latter will also decrease when the electrostatic well disappears and dehydration begins to control the current. At large voltage, the current becomes less sensitive to strain because the applied bias dominates over the energy landscape of the pore, self-consistently washing out relevant features---ones that are contributing to resistance---of the landscape. As voltage increases further still, the current saturates at a smaller strain where the relevant free energy features are commensurate with the voltage drop (\textbf{bottom panel}). The conductance versus strain shows that for $\qO=-0.24\,e$ there is an intricate interplay of voltage and strain, indicating that the variation of free energy features with these two parameters is playing a defining role. At larger strain (greater than about 6\%), the conductance tends to increase with the voltage (i.e., superlinear behavior). This is a telltale sign of an activated process, where the voltage helps overcome an overall barrier, but does not yet wash it out. In this particular case, this is due to a reduction in electrostatic compensation of dehydration as strain pulls away the counteracting negatively charged oxygen atoms. For $\qO=-0.54\,e$, the conductance increases with the voltage at smaller strain (superlinear behavior) and decreases with  voltage at larger strain (sublinear behavior). The superlinear behavior indicates barrier-limited transport and sublinear behavior diffusion-limited. The error bars are plus/minus one SE from five parallel runs. Connecting lines are a guide to the eye only. Purple shaded regions in the upper panels approximately delineate the region where bulk limitations control the current. \label{fig:IS}
 }
\end{figure}

\begin{figure}[H]
\centering
\includegraphics[width=7 cm]{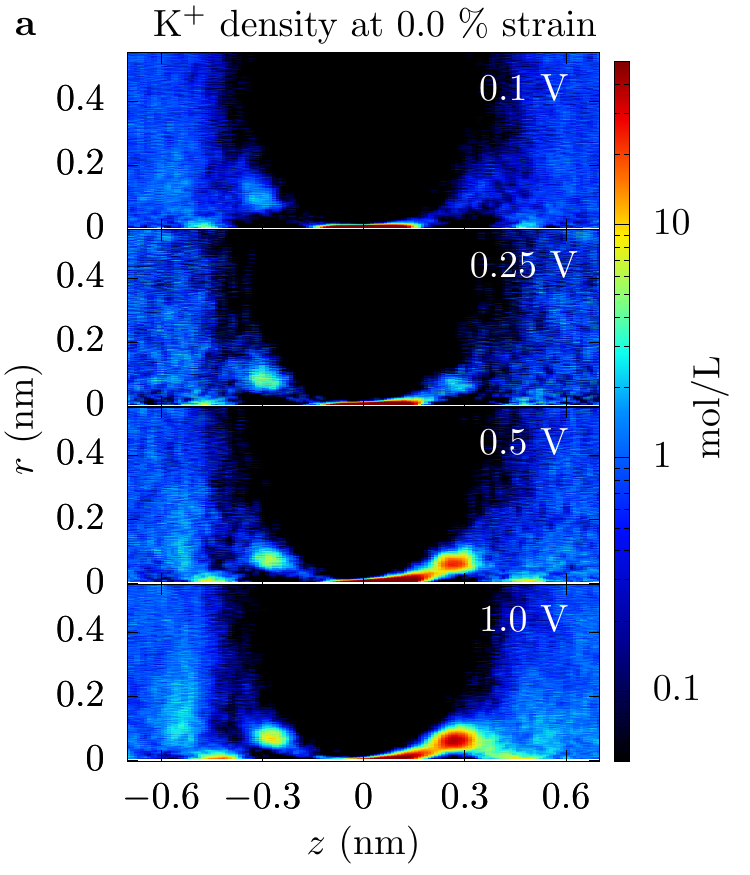}
\includegraphics[width=7 cm]{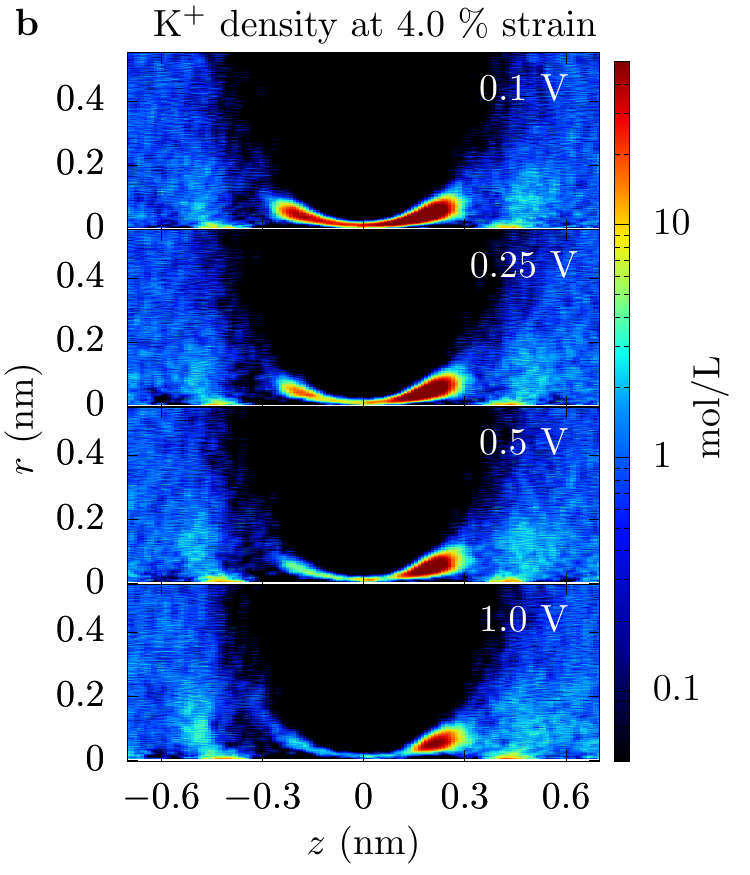}
\caption{{{\bf  Voltage dependence of the ion concentration.}} 
 Concentration of potassium ions near a graphene crown ether pore with $\qO=-0.54\,e$ at ({\bf a}) 0\% and ({\bf b}) 4\% strain for various voltages. For 4\% strain, where we see diffusion limitations of the ionic current, we also see the depletion of ions in the pore as the voltage increases. See the Supplemental Material (SM) for additional plots with different parameter values. \label{fig:CK}}
\end{figure}

{{\textbf{Ionic current through a graphene crown ether pore:}} 
 } Figure~\ref{fig:IS} shows the potassium current, $I_\mr{K}$, through the graphene crown ether pore versus strain at various voltages. Only potassium ions contribute to the total current, as the negatively charged pore edge does not allow any chloride ions to translocate (on the timescale of the simulations). We also plot the conductance of potassium ions, $G_\mr{K}$, in order to demonstrate the non-Ohmic behavior of ionic current. At low voltage, the current increases by several fold for a minute strain---a couple percent strain changes the conductance by a couple hundred percent. This dramatic amplification is an example of colossal ionic mechano-conductance~\cite{sahu2019optimal}. The current eventually maximizes around 3\% strain and either decreases (for $\qO=-0.24\,e$ at small voltage) or plateaus (for all other cases).
Since the pore size does not change substantially with a small strain, the colossal change in the ionic conductance is the result of a modification of the translocation barriers. The translocation landscape veers toward barrierless transport as strain tunes transport to its optimum~\cite{sahu2019optimal}. Furthermore, the change in conductance with voltage displays non-Ohmic behavior. In some regimes, such as the colossal mechano-conductance, the conductance increases with voltage, indicating an activated process. In other regimes, the conductance decreases with voltage, indicating the diffusion-limited process. The depletion of charge carriers in the pore, see Figure~\ref{fig:CK}, is also consistent with the decrease in conductance and points to a diffusion-limited regime.

{\textbf{Translocation barriers in sub-nanoscale pores:}} 
The single-ion energetics of transport through functionalized sub-nanoscale pores may be approximately expressed as
\begin{align}\label{eq:energy}
\Delta F_\nu & \approx \sum_i \eta_i f_{i\nu } E_{i\nu} + \sum_{\nu^\prime} \frac{ q_\nu q_{\nu^\prime} n_{\nu^\prime}}{4 \pi \epsilon_0 \epsilon\, r_{\nu \nu^\prime}},
\end{align}
where $f_{i\nu}$ and $ E_{i\nu}$ are the fractional dehydration and energy of $i^\mr{th}$ hydration layer for ion $\nu$, $q_{\nu^\prime}$'s and $n_{\nu^\prime}$ are the charge and number of atom species $\nu^\prime$ in the pore, and $\epsilon_0$ is the vacuum permittivity. The parameter $\eta_i$ is an $\mathcal{O}(1)$ factor to account for the increased binding of water molecules with the ion as dehydration increases~\cite{sahu2017dehydration}, essentially giving the non-linear response of the hydration energy to the removal of water molecules. The relative permittivity of water, $\epsilon$, under nanoscale confinement is significantly smaller than the bulk value and depends on atomic details~\cite{fumagalli2018anomalously, gibby2018role, sahu2019optimal}. Specifically, in the case here, when there are not intervening water molecules between the ion and charged groups in the pore, the dielectric constant is around 4~\cite{sahu2019optimal} and the electrostatic interaction is very large. The small dielectric constant and short distances involved give rise to the large electromechanical susceptibility of ions within the pore. The fractional dehydration $f_{\nu 1}$ also changes with the position of ion~\cite{sahu2017dehydration, sahu2017ionic, sahu2019optimal} and can be estimated with geometric arguments~\cite{zwolak2009quantized,zwolak2010dehydration,sahu2017dehydration,sahu2017ionic}. As an ion approaches the pore, the free-energy change will remain small ($<\kt$) even at 1~nm distance from the pore, because the ion is still fully hydrated. The electrostatic interaction between the fully hydrated ion and the pore charge is weak due to the dielectric screening of the solution. However, when the ion is about 0.5~nm from the pore, it starts to dehydrate (initially in the second hydration shell and then in the first), and consequently the dehydration energy increases sharply. Simultaneously, the electrostatic energy also rises rapidly since the ion will be significantly closer to the negatively charged oxygen atoms compared to the positively charged carbons, and the effective dielectric constant of water at this distance will be strongly diminished due to the removal of intervening molecules.

\begin{figure}[H]
\centering
\includegraphics[width=14 cm]{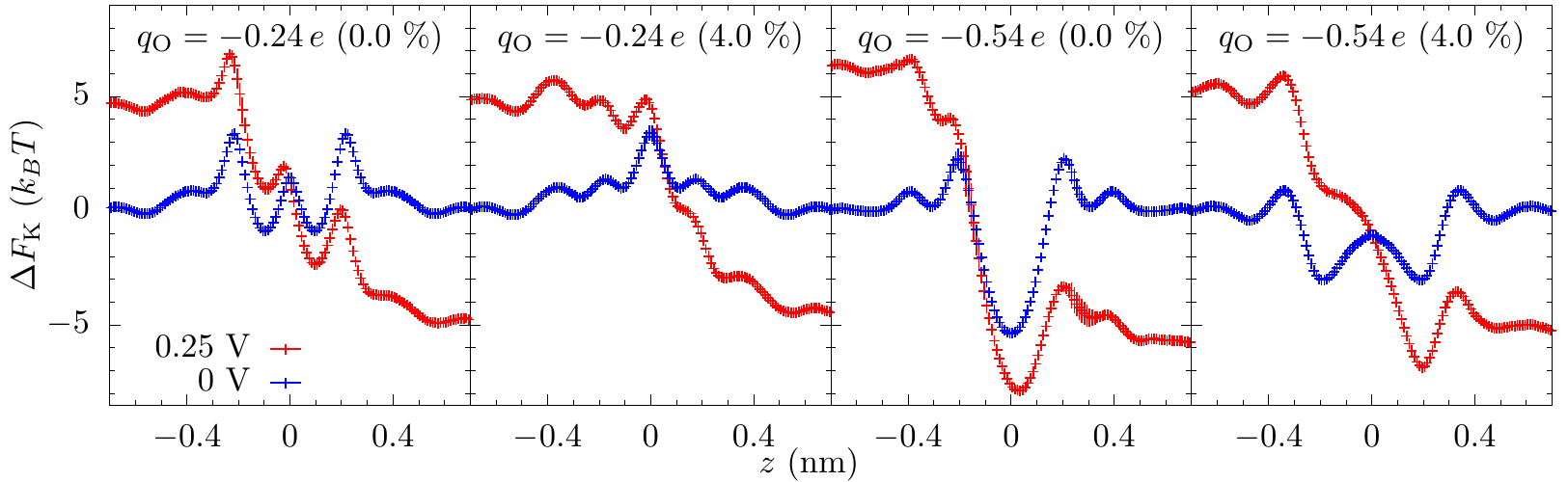}
\caption{{{\bf Equilibrium and voltage-dependent landscape for ion transport.}} 
 The free-energy profile of K$^+$ going through a graphene crown ether pore at 0\% and 4\% strain for equilibrium and non-equilibrium ($\Va=0.25$~V) cases. The charge of the oxygen atoms of the crown ether is either $-0.24\,e$ or $-0.54\,e$  (and adjacent carbon atoms have half this charge). The potential wells and barriers are mainly the result of competition between the electrostatic attraction and the dehydration. The applied voltage reduces the features (barriers and wells) in the energy landscape but some sharp features still remain, either due to the barriers' size or due to their {\em irrelevance}, a term we use operationally, see the main text and Figure~\ref{fig:grad-V}. Irrelevance of barriers occurs since the influence of a barrier on the current is both a kinetic and thermodynamic effect, and other bottlenecks (e.g., diffusion limitations) can exist, i.e., these are the relevant processes at a given voltage and strain. In particular, for 4\% strain and $\qO=-0.54\,e$, the supply of ions from bulk has a much larger influence on the ionic transport than the dissociation from the well at 0.2~nm. Hence, that barrier remains roughly unchanged. Error bars are plus/minus one SE from five parallel runs. \label{fig:pmf-V}}
\end{figure}

Equation~(\ref{eq:energy}) gives these qualitative features of the energy landscape and helps to understand why electrostatics can play such a strong role even in high salt solutions. Still, we use all-atom molecular dynamics (MD) simulation for the calculation of quantitative landscape---see the Methods and later discussion about an additional entropic barrier to move into the ABF constriction. The free-energy profiles from MD are shown in Figure~\ref{fig:pmf-V}. Since we are driving the ionic current through a nanopore by an external voltage, we also calculate the energy landscape of ion transport with an applied bias. The equilibrium free-energy barrier alone does not fully represent the energy landscape of ion transport, especially when the applied bias is large compared to the features in the free energy. We note, of course, that even the energy landscape with the bias does not fully capture the current due to kinetic prefactors and averaging~effects. 

The equilibrium free-energy profiles (blue lines in Figure~\ref{fig:pmf-V}) exhibit a potential barrier for $\qO=-0.24\,e$ and potential well for $\qO=-0.54\,e$ at the center of the pore in the unstrained membrane. In the former, the electrostatic energy (between the ion and the pore charges) is less than the dehydration barrier, while the opposite is true for the latter. Additionally, there can be small potential wells just outside the pore where the ion maintains a larger hydration yet stays close to the negatively charged oxygen atoms of the pore. The energy landscape changes markedly with strain, which is primarily due to the change in the electrostatic interactions within the pore and dehydration outside of the pore. An increase in the pore size---by  picometers---due to the strain causes the attractive electrostatic energy to decrease rapidly. The dehydration energy penalty in the pore also decreases with strain but, for small strain, it does not change as rapidly as the electrostatic energy. Consequently, there is a net increase in the energy of the ion at the center of the pore. As a result, the barrier in $\qO=-0.24\,e$ increases, and the potential well in $\qO=-0.54\,e$ flattens and then disappears at large strain (a dehydration-based barrier does appear in the middle of the $\qO=-0.54\,e$ pore, a feature which is already present in the $\qO=-0.24\,e$ pore at 0\% strain due to the lower electrostatic compensation). In contrast, the effect of the strain on the free-energy outside the pore is in the opposite direction. The barrier outside the pore decreases with strain as ion can hydrate better with reduced hindrance from the pore oxygen atoms. The electrostatic energy, however, does not change as rapidly as in the center of the pore. The basic mechanism behind these large changes in free energies is that at the 0.1~nm to 0.5~nm scale; picometer changes in atomic configuration result in large changes in electrostatic and dehydration energies~\cite{sahu2019optimal}. Dielectric screening (from the solution), in particular, is not that effective at this length scale. 

The free-energy landscape explains many of the features seen in the ionic current versus strain, Figure~\ref{fig:IS}. At small voltages, the current changes significantly with strain because of the change in the energy landscape of the pore. For $\qO=-0.24\,e$, the entrance barrier, just outside the pore, initially decreases with strain and the current increases rapidly. Eventually, the increase in the energy at the center of the pore will negate the decrease in the outer barrier, and the current subsequently decreases, thus giving a turnover behavior with the optimal current around 3\% strain. At very large strains, ion hydration will increase in the pore, and thus the energy barrier at the center will start to disappear. For $\qO=-0.54\,e$, the potential well at the center of the pore becomes shallow with strain, making it easier for ions to dissociate from the pore and contribute to the increase in the current. A common principle for the colossal mechano-conductance change is that the free energy veers toward a barrierless landscape for both these example pore environments. At intermediate strains (4\% to 10\%), while strain does influence barriers, bulk limitations have kicked in and the barrier change will not be manifest in the current versus strain. Even at small strain, the current will become flat if the applied voltage is large enough, since the larger voltage can wash out larger free energy features. There are, however, irrelevant free energy features---ones that are not rate limiting---that remain even as voltage increases, for which we introduce discrete-barrier and one-way rate analyses below that help identify relevant and irrelevant features. The evolution of features under strain can also suggest their relevance (i.e., if the current is constant versus strain, yet a large feature disappears, that feature is likely---but not guaranteed, since other factors can conspire together---to \mbox{be irrelevant). }

\newpage
To elucidate the effect of the applied bias on the energy landscape, and hence the current, we calculated the free-energy of the potassium ion in the presence of an external voltage ($\Va=0.25$ V). Figure~\ref{fig:pmf-V} shows that the applied voltage raises the potential on one side of the membrane and decreases on the opposite side and the overall potential roughly drops over $|z|<0.5$~nm (we also see this drop in the calculation of the electrostatic potential). Note that although the graphene is only 0.3~nm thick, the double layer of cations and anions on the opposite side of the membrane will be separated by a distance of about 1~nm due to their hydrated radii. Nonetheless, even a small voltage will result in a large electric field in the pore which can suppress the energetic features. Yet, some sharp features have a spatial variation larger than the applied field and are still prominent in the free energy landscape with an applied voltage. 

\begin{figure}[H]
\centering
  \includegraphics[width=14 cm]{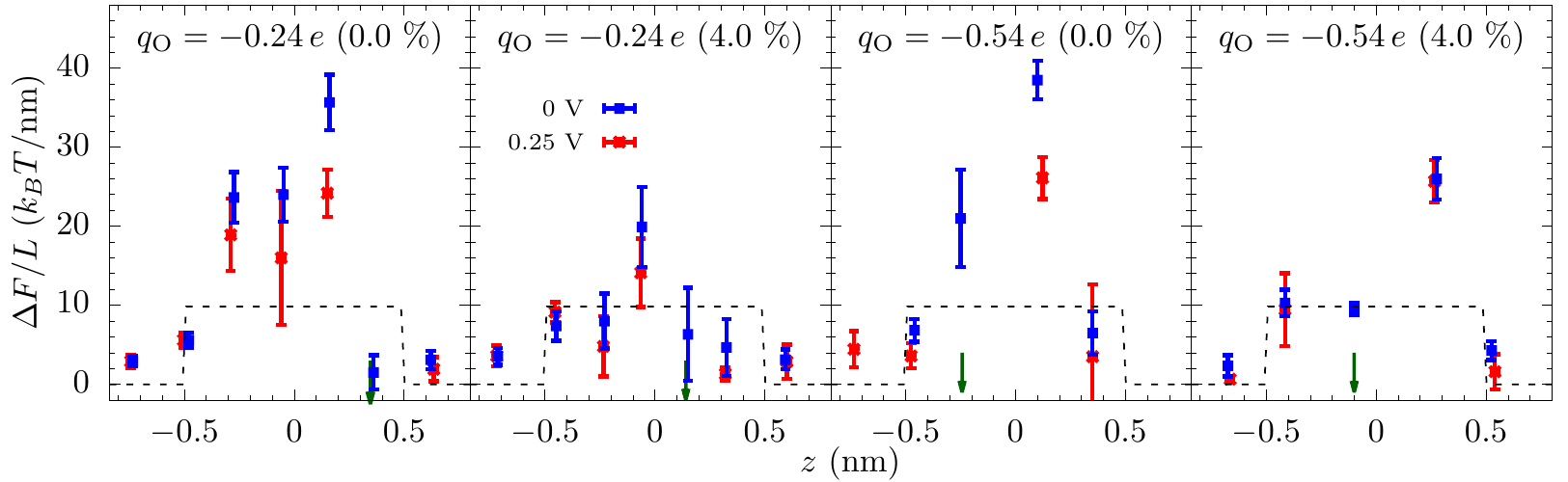}
\caption{{{\bf Discrete gradients and the renormalization of energy barriers.}} 
 The discrete energy gradients encountered by a K$^+$ ion in graphene crown ether pores at 0\% and 4\% strain for 0~V (blue) and 0.25~V (red) applied bias. The gradients are between each local minimum and the next maximum in the positive $z$-direction in Figure~\ref{fig:pmf-V} and we plot them against the mean position of the barrier. The gradient from the applied bias ($\Va=0.25$ V), ideally about 10 $\kt$/nm between $z=\pm 0.5$ nm (shown with black, dashed line), reduces the translocation barrier, completely eliminating it in some cases (shown with a green arrow). To the first approximation (in particular, ignoring the self-consistent development of the potential drop), the sharp features that are larger than the ideal electric field will have a significant influence on the ionic current, as they are still present (though reduced) when the voltage is applied. This ideal behavior is approximately occurring in the $q_O=-0.24\,e$ pore, as for both values of strain shown the gradients within the membrane region are being collectively diminished. For $q_O=-0.54\,e$, more complex behavior is occurring, with some features changing more than others. Examining the change in discrete energy gradients upon application of a voltage gives a clear indication of the presence of irrelevant features---these barriers do not change, as they are not rate limiting and do not create a self-consistent potential drop around themselves. The errors are due to the uncertainties in both the position and the magnitude of minima and maxima.\label{fig:grad-V} }
\end{figure}

To capture how the features in the equilibrium free-energy profile change with applied voltage, we plot in Figure~\ref{fig:grad-V} the discrete gradients from each energy minimum, $i$, to the next maximum in positive $z$-direction, i.e., $(\Delta F^i_\text{max}-\Delta F^i_\text{min})/(z^i_\text{max}-z^i_\text{min})$. The gradient of applied voltage, which is in the opposite direction to these gradients, reduces the barrier to transport (we do not plot the gradients in negative $z$-direction, which assist rather than hinder the ion translocation). Figure~\ref{fig:grad-V} shows that some of the gradients are larger than the electric field from the applied bias ($\Va=0.25$ V), and thus these barriers are still present in the energy landscape with applied bias. More importantly, though, the examination of how these discrete gradients change with voltage enables one to identify rather large features that remain unchanged at finite voltages, such as the well at 0.2 nm for the $q_O=-0.54\,e$ and 4\% strain case. This well (and associated barrier) is not a limiting factor in transport at this strain and thus the applied bias does not self-consistently remove it. This type of plot (and a related plot we will examine later) give a clear depiction of what features are influencing transport, including indirectly the influence of kinetic prefactors. We further note that the largest gradient  for the unstrained pore at $q_O=-0.54\,e$ is about 40~$\kt$/nm and thus will require $\Va\approx 1$~V to effectively wash it out, which is reflected in Figure~\ref{fig:IS}. Once the applied voltage produces local fields larger than the relevant discrete gradients, the ionic current will have little dependence on the equilibrium landscape of the pore, which explains the saturation of the ionic current across all values of strain for large voltages, as we see in Figure~\ref{fig:IS}. Saturation at smaller voltage is a combination of this same washing out plus the presence of irrelevant features due to high kinetic rates (compared to other rates, such as diffusion and entrance-side feeding; see the Supplemental Material (SM) for additional plots of the equilibrium and non-equilibrium free energy barriers). 

Some features in the energy-landscape, though, are beyond 0.5 nm from the pore, albeit they are small. These features can survive to large applied voltages. Thus, while they matter little for smaller voltages, they eventually can become important when their energy- and kinetic-scales are commensurate with the other renormalized features. Thus, ions will eventually have to overcome additional entrance barriers. These barriers will directly affect the rate at which ions can enter and exit the pore and thus influence the saturation current through the pore. Conversely, the barrier on the exit side, though significant, has a smaller influence due to the larger dissociation rate, which we will discuss later when examining the interpretation of the rate constants within the model. 

We note that since the K$^+$ ion is confined to a cylindrical region during the ABF calculations, the free energy we present does not include the entropic, `constriction' barrier to move an ion from bulk to the ABF cylindrical constriction of radius  $r_{\mathrm{ABF}}=0.28$~nm, and vice versa (on the exit side). The ABF constriction allows other ions (both coions and counterions) to be in the volume. The ratio of accessible states is thus approximately $\Omega_{\mathrm{ABF}}=\pi r_{\mathrm{ABF}}^2 l/l^3$, where $l$ is the typical distance between co-ions in bulk ($\approx$1.2~nm at 1~mol/L KCl). Thus, the contribution to the free energy of this constriction penalty is $-\kt \ln{\Omega_{\mathrm{ABF}}}\approx 1.7\,\kt$. From within the ABF constriction, the entropic penalty to then go into the pore is included within the ABF calculation. For comparison, this contribution can also be estimated as follows: The geometric pore radius is $r_p=0.137$~nm, taken as the pore center to oxygen center, 0.29~nm, minus oxygen's van der Waals radius, 0.152~nm. A typical approach to estimate the entropic penalty is the formula, $-\kt \ln{\left(1-(r_{\mathrm{K}^+}/\a)^2\right)}$. However, $r_{\mathrm{K}^+} \approx \a$ (see, e.g., Ref.~\cite{mahler2012study} for ionic sizes) and this approach will lead to large errors and, in fact, does not include important physical processes, such as the movement of oxygen atoms at the pore rim. A better approach is to estimate the entropy from the actual trajectories of ions going through the pore. Potassium ions cross the pore within a radius of about $r_c\approx0.02$~nm from the origin. Assuming that the ions are not localized in a well, but still are locally in equilibrium, the entropic penalty is approximately $-\kt \ln \left( r_c^2/r_{\mathrm{ABF}}^2 \right) \approx 5~\kt$. The presence of a well of size $l_W \approx 0.3$~nm in some cases gives an additional contribution $-\kt \ln \left( l_W/l \right) \approx 1~\kt$ to $2~\kt$.

{\textbf{Radius of the pore: }} 
Before moving forward, we address an issue that permeates the whole field of transport in sub-nanoscale pores and is apparent in the proceeding paragraph---that of the pore radius. For smooth, uncharged pores, the radius or open area (when not circular) can be rigorously defined with all-atom simulation: One samples the trajectories of ion crossings and takes a weighted average of discrete area elements (see Ref.~\cite{sahu2018maxwell}, where the current density was roughly uniform, enabling a direct and intuitive treatment). However, when pore charge is present or the pore has structure, whether steric or energetic, along its length, there is clearly no simple answer for pore radius. The effective radius that defines access resistance, for instance, will not be the same as the geometric radius of the pore mouth. This is easy to see when charge is present at the pore mouth, since the effective opening within a continuum approximation will increase by about a Debye length due to electrostatic attraction of counterions~\cite{sahu2019colloquium}. The fact that effective sites are present will change this picture further. For instance, there are association-side sites that form a staging area from around $z=-0.4$ nm to $-0.3$ nm with a spread $r_s$ of about 0.1 nm, which is related to $\a$ but can be influenced by other factors (their numerical values here are the same). It is this region that has to be ``accessed.''

Moreover, if the pore has a conical shape (e.g., even for this graphene crown ether pore, ions seem to follow a coarse canonical shape, see Figure~\ref{fig:CK}), what radius is relevant to defining the ``open area'' of the pore, especially when energetic features are present? When variation in size or energy is large on the scale of inter-ion separation and the ion mean-free path, this issue can be handled simply by assuming local equilibrium and appropriately averaging. The graphene pore examined here, as well as other pores in 2D membranes and biological channels, do not have such a simple separation of scales. Fortunately, here, the important length scales that define size fall within the range $r_c \approx 0.02$~nm (spread of trajectories of ion crossings) to $r_s \approx 0.1$~nm (spread of association-side sites) to $r_p=0.137$~nm (geometric radius) to $\lambda_D \approx 0.3$~nm (Debye length). We will take the effective pore radius as $\a \approx 0.1$ nm. This value is in the middle of this range and thus, except for a few particular quantities such as the entropic barrier, it gives a reasonable starting point for estimating values of different pore characteristics.

\begin{figure}[H]
\centering
\includegraphics[width=0.9\textwidth]{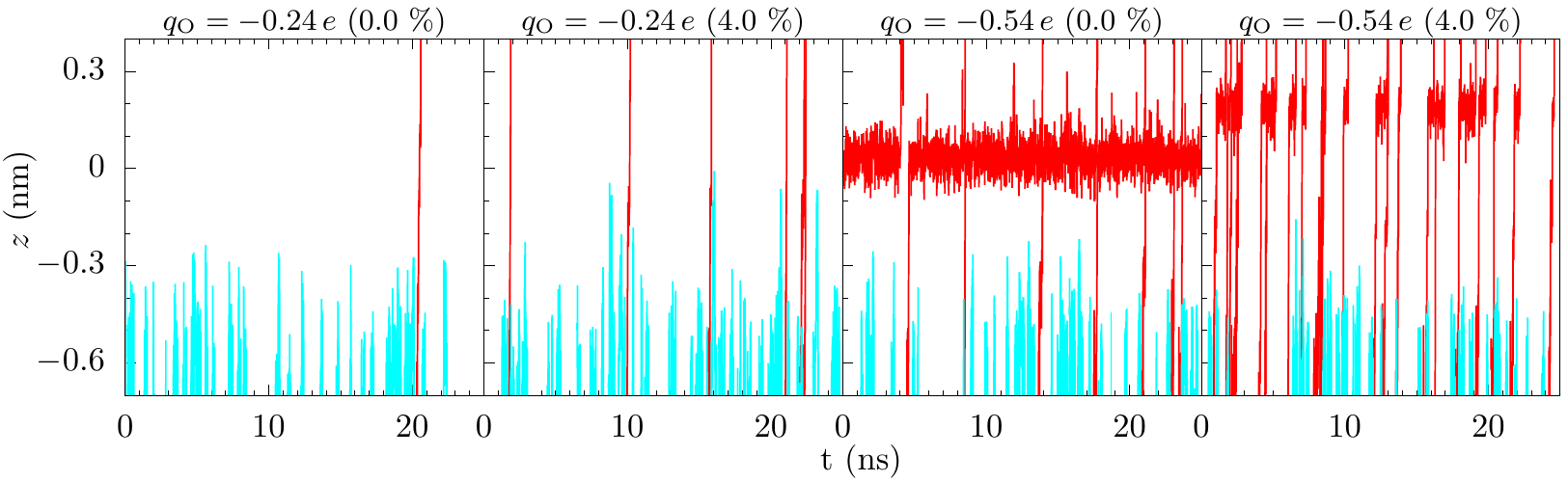}
\includegraphics[width=0.9\textwidth]{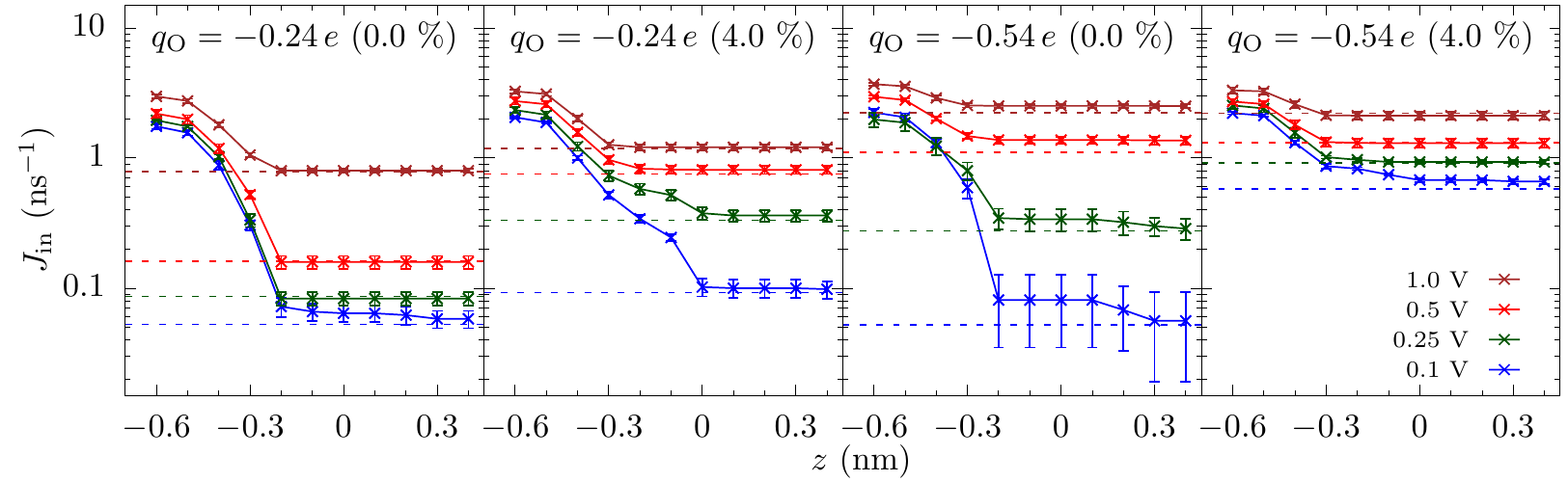}
\caption{{{\bf Translocation events and one-way rates.}} 
 (\textbf{Top panel}) Time trace of the $z$-position of potassium ions that translocate (red) through the graphene crown ether pore and reflect (cyan) after coming within 0.6 nm of the center of the pore. For $\qO=-0.24\,e$, ions cross the pore very quickly and the association rate is the primary determinant of the current. In the unstrained pore with $\qO=-0.54\,e$, the ions spend a significant time in the pore, and thus the dissociation rate determines the current. (\textbf{Bottom panel}) The inward rate of K$^+$ ions versus $z$-distance at different applied voltages. The dashed horizontal lines gives the net rate. For small voltage, $J_\text{in}$ near the pore is much smaller than the bulk diffusion rate, and thus the current is limited by the barriers to transport. Error bars are plus/minus one SE. Connecting lines are a guide to the eye only. \label{fig:rate}}
\end{figure}

{\textbf{Incoming rates: }} 
Before discussing the modeling of these pores, we first introduce a simple tool to further assess the influence of different energetic features. Figure~\ref{fig:K-traj} shows the trajectories of K$^+$ ions moving toward the graphene crown ether pore. The trajectories of ions that eventually translocate through the pore are shown with a red line, and others are shown in cyan lines. Only a few non-translocating trajectories go into the range of $z=-0.3$ nm. This becomes more apparent by plotting the trajectories near the pore (within the radial distance of 0.6 nm from the center of the pore) versus $z$ and time, as seen in Figure~\ref{fig:rate} upper panel. Information regarding the rejection of ions would thus be helpful. In Figure~\ref{fig:rate}, we thus also plot the incoming rate $J_\text{in}$ of ions crossing a $z$-plane versus the $z$-distance at various applied voltage. 
Initially, $J_\text{in}$ drops rapidly with $z$, as ions have to go through a diffusion constriction and also get reflected by the entrance barrier. At a certain location, $J_\text{in}$ becomes flat, indicating all ions that made it to that distance will complete the translocation. For $\qO=-0.24\,e$, for example, the rate drop sharply going from $z=-0.5$ nm to $z=-0.2$ nm in the unstrained pore due to the presence of an occupation barrier. The rate then becomes flat, as ions cannot go back (we note that we do see some ion crossing events that go backward, up the potential gradient. These are few and far between, but the small gap in Figure~\ref{fig:rate} for some cases quantifies this magnitude of these events). For 4\% strain, the rate continues to drop until $z=0$, as there is a large barrier at the center of the pore. 

Similar observations can be made for $\qO=-0.54\,e$. For the unstrained pore at small voltage, we see a large drop in the incoming rate between $z=-0.5$ nm and $z=-0.2$ nm due to the repulsion from the ion already in the pore. There is a smaller drop due to dissociation of the ion from the pore. Importantly, both of these drops are due to dissociation, with the former due to a blockade (many-body) effect and the latter being actual ion dissociation. We note that many-body and single-ion effects can be unraveled by comparing the free energies at finite concentration to the free energy of a single ion pair in solution~\cite{sahu2019optimal}, which shows that the satellite barriers for the unstrained, $\qO=-0.54\,e$ pore are due to the presence of an ion in the pore. For the 4\% strain (and for the unstrained pore at larger voltage), the drop in the rate is small and it essentially saturates at $z=-0.3$ nm. This means that ions do not feel a significant barrier going through the pore and the total current is only limited by the rate at which ions arrive at the mouth of the pore. As with the discrete barrier gradients, the plot of the one way ion rate allows for the identification of what features matter. For $\qO=-0.54\,e$ at strain at about 4\% and above, the reduction in ion flux at the entrance side is due to the diffusion constriction and entrance barriers. These incoming ion rate plots thus provide both qualitative and quantitative information. We will use this to motivate the modeling choices below (specifically, the use of a staging site and the assumption of one-way current flow in the pore).

{\textbf{Reaction rate model:}} 
The 18-crown-6 pore in graphene can only fit a single ion at a time. It is thus intuitive to analyze the ionic transport process using rate theory~\cite{lauger1973ion}: Ions arrive at the pore at a certain rate and depart at a certain rate, which together provide the ionic current. 

The simplest case would be to assume a single site and that ions only move in one direction. The latter takes into account that the bias is sufficiently large that ions cannot move backward, up the potential gradient (this is a reasonable assumption for the voltages in this work, as we saw above, but cannot correctly reproduce equilibrium conditions). In that case, the ionic current through the pore is given as $I/q=\left(1/k_a + 1/k_d \right)^{-1}$, where $k_a$ ($k_d$) is the rate constant for association (dissociation) of ions into (from) the pore. In terms of the site occupancy (equivalently, probability of being occupied), $I/q=k_a\,(1-P)=k_d\,P$ and $P=k_a/(k_a+k_d)$. These latter equations make it clear that, with a strain independent association rate $k_a$, the current will linearly depend on occupancy and thus cannot plateau, as seen in Figure~\ref{fig:IS}, until $P$ is effectively zero. This limit, $k_a\ll k_d$, gives $I= q\,k_a $, in which case the ionic current is fully determined by the incoming rate. 

We will see that $P$ is still substantial on some of the plateau. Thus, while the fit to a single-site model is reasonable when allowing $k_a$ to have some voltage dependence (i.e., $k_a=k_{a0}+\kappa_a V$), the model is not qualitatively consistent with the data, as the model current still increases when the actual current has leveled off. This assessment of the single-site model is the same regardless of whether only one way motion is assumed or not: Allowing fluctuations in and out of the pore on both sides of the membrane still gives a linear dependence on $P$ with a similar coefficient. 

Instead, we examine a three-site model, despite the fact that the channel is atomically thin. The data in Figure~\ref{fig:CK} show that there are multiple localized regions of enhanced K$^+$ density. On the association side (left side of the figure), there are candidate sites---staging sites---at about $-0.3$ nm and about $-0.4$ nm (offset from each other also in the radial direction), and similarly on the dissociation side. That is, there are 4 or 5 candidate sites in the parameter regimes of that figure (ambiguity results from the fact that the candidate sites on the dissociation side are not fully disconnected---there is a non-negligible probability to find an ion in between some locations). These sites are due to ripples in the free energy, which extend outside the pore, as discussed above and seen in Figure~\ref{fig:pmf-V}. Due to the proximity of the association-side staging sites, we will assume they are the same and employ a three site model. Moreover, the one-way rate data in Figure~\ref{fig:rate} supports this view of the pore, as well as the assumption that current (mostly) flows in one direction at the pore binding site. 

The kinetic equations for the three site system are 
\begin{eqnarray}
\dot{P}_1 & = & k_b (1-P_1) - k_b^\prime P_1 - k_a P_1 (1-P_2) \label{eq:P1} \\
\dot{P}_2 & = & k_a P_1 (1-P_2) - k_d P_2 (1-P_3) \label{eq:P2} \\
\dot{P}_3 & = & k_d P_2 (1-P_3) + k_{b^\prime} P_3 - k_{b^\prime}^\prime (1-P_3) , \label{eq:P3}
\end{eqnarray}
where $P_i$ is the occupancy of the site $i=$1, 2, and 3, $k_b$ ($k_b^\prime$) is the incoming (outgoing) rate from bulk on the association side, and  $k_{b^\prime}$ ($k_{b^\prime}^\prime$) the dissociation side. Again, the set of equations assume only one way motion into (the association side), and out of (the dissociation side), the internal pore site $i=2$. Backward fluctuations can easily be included, but this adds extra parameters to be fitted and will only influence the fit in a minor way. We will apply this model only to the behavior of the $q_O=-0.54 \, e$ pore, since the $q_O=-0.24 \, e$ pore has more intricate behavior that would ultimately require association rates that are strain-dependent, i.e., that depend on the variation of the free energy landscape. We have discussed the $q_O=-0.24\,e$ pore extensively already in Ref.~\cite{sahu2019optimal}, including the origin and scale of the free energy variation. We only note here that, as seen in {Figure~\ref{fig:IS}}
, the 0.5~V and 1~V biases for $q_O=-0.24\,e$ also give an entrance-limited region. The magnitude of the currents in this region are lower than $q_O=-0.54\,e$ by only an order one factor for the same voltages and strain. The similarities in current are expected for bulk-limited behavior. The fact that they are lower by a small amount is likely due to the increased capture effectiveness of the higher charge pore. The specific estimates for parameters will thus apply in this case, albeit with some small modifications of effective radii and rates.

Even with the assumptions regarding one-way rates at the $i=2$ site and the symmetry of bulk rates, there are a number of parameters. Instead of direct fitting, we can employ main pore site ($P_2$) occupancy data from MD to reduce the number of parameters and see if a consistent model results. Considering only Equation~\eqref{eq:P1} and setting $\dot{P}_1=0$ yields the occupancy of the first site
\begin{equation} \label{eq:P1val}
    P_1=\frac{k_b}{k_b+k_b^\prime+k_a(1-P_2)}.
\end{equation}
This site and the third site are the least well-defined, and thus eliminating them from the expressions is key to reducing mathematical and computational acrobatics in defining and fitting the quantities in the model. The particle current is given by the last term in Equation~\eqref{eq:P1} (or, equivalently in the steady state, the sum of the first two terms), 
\begin{eqnarray}
I/q & = & k_a P_1 (1-P_2)= \frac{k_b k_a (1-P_2)}{k_b+k_b^\prime+k_a(1-P_2)} \\
& = & \left( \frac{1}{k_b} + \frac{1}{\tilde{k}_a (1-P_2)} \right)^{-1}, \label{eq:model}
\end{eqnarray}
where $\tilde{k}_a=k_a P_1^{\mathrm{eq}}$ is the effective association rate and $P_1^{\mathrm{eq}}=k_b/(k_b+k_b^\prime)$ is the equilibrium density of site 1 in the absence of its connection to the main pore site (in this absence, we can examine equilibrium of $P_1$). We do not have to separately determine or fit $P_1^{\mathrm{eq}}$, since we can examine solely $\tilde{k}_a$ for association and only $k_b$ to give the influence of bulk. Note that Equation~\eqref{eq:model} has made no assumptions regarding the relative magnitude of the dissociation rate, or, for that matter, the influence of any of the factors that appear in Equations~\eqref{eq:P2} and~\eqref{eq:P3}, other than the $k_a P_1 (1-P_2)$ term common with Equation~\eqref{eq:P1}. Thus, the model can be thought of as just Equation~\eqref{eq:P1}, which has only the assumption that there is a negligible backward rate from the site 2 to site 1, which as we have seen is justified for much of the parameter ranges examined for $q_O=-0.54\,e$. The form of the bulk rates on the dissociation side and the lack of backward processes on that side is thus inconsequential. Moreover, whether the model is two or three sites is also irrelevant due to our approach. The inclusion of $(1-P_2)$ in the model, which will be directly extracted from MD, captures the influence of all potential processes on the dissociation side, whether included in Equations~\eqref{eq:P2}~and~\eqref{eq:P3} or not.

\begin{figure}[H]
\centering
    \includegraphics{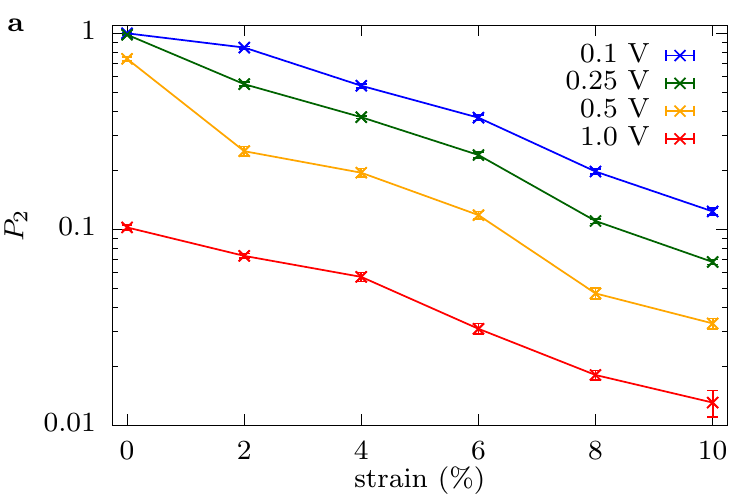}
    \includegraphics{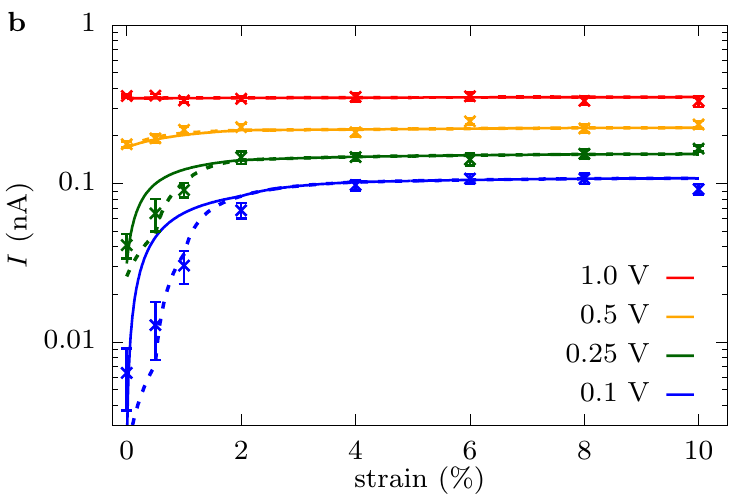}
\caption{{{\bf Pore occupancy versus strain and model fit to the current for the $q_O=-0.54\,e$ pore.}} 
({\bf a}).~Pore occupancy versus strain for the four voltages indicated. The occupancy is decreasing exponentially with strain and voltage, with some additional, minor features and an apparent threshold behavior with voltage at zero strain. These data are reproduced in the SM along with data for the $q_O=-0.24\,e$ pore. ({\bf b}). Current versus strain at the four voltages labeled, along with the model. The latter was fitted using current-voltage and $P_2$ data at 0\% to 10\% strain at 2\% increments (i.e., the $P_2$ data shown in {\bf a}). The continuous model plot is found by linearly interpolating the $P_2$ data. The model is very good when accounting for diffusion and access limitations. When using the interpolated $P_2$ data from 2\% increments (solid line), there is some deviation at 0.5\% and 1\% strain. However, using the interpolated $P_2$ data including those two additional points (dashed line) shows that the issue is that $P_2$ has features not captured by interpolation at 2\% increments (see the SM for the additional $P_2$ data). The $R^2$ and adjusted $R^2$ for the fit are 0.998 and 0.997, respectively (for data from 2\% increments). The step-like features are solely due to employing (linear) interpolation to create a continuous curve. Error bars are plus/minus one SE. Connecting lines are a guide to the eye only. 
\label{fig:model} }
\end{figure}

Figure~\ref{fig:model} shows the occupancy of the pore, $P_2$, and the model results overlaid with the current data. Note that we only fit the model for select points (the one for which occupancy data is shown). Since the conductance depends on voltage, we let $k_b=k_{b0}+\kappa_b V$, which together with $\tilde{k}_a$ gives a three parameter fit. The resulting fit parameters are $k_{b0}=(0.50 \pm 0.03)\times 10^9$~ion/s, $\kappa_b=(2.2 \pm 0.2) \times 10^9$~ions/(V$\cdot$s), and $\tilde{k}_a=(1.2 \pm 0.3) \times 10^{10}$~ion/s, with uncertainties given by the standard error of the fit. We will discuss these parameters shortly, including their agreement with back-of-the-envelope estimates, as well as providing a quantification of diffusion limitations versus drift-supplied ions. 

When the model and data are viewed in tandem, the physical behavior is apparent. When the current plateaus versus strain, it is due to combined diffusion and entrance/access limitations, for which without some component of the latter, the current would not increase substantially with voltage (the voltage could only decrease local ion density, increasing and eventually saturating the diffusive contribution in the process). For smaller strains and voltages, the current is dominated by the $\tilde{k}_a (1-P_2)$ component. That is, the current is dictated by a many-body effect: localization of a K$^+$ ion prevents current flow until that ion dissociates, in which case an effective particle current of $\tilde{k}_a$ flows while the pore is empty (i.e., in more concrete terms, this regime can be thought of as a current of zero flowing, while the $i=2$ site is occupied and $\tilde{k}_a$ otherwise, giving $I/q=0 \cdot P_2 + \tilde{k}_a (1-P_2)$ and considering $P_2$, which is between 0 and 1, to be a probability). The many-body nature of transport in this regime is further supported by a decreasing ionic current versus concentration, which shows the saturating nature of the process; see the concentration figures in the SM.

The expression in Equation~\eqref{eq:model} quantitatively captures the current versus strain and voltage behavior for most of the data. Where it gives the least fidelity to the full simulation result (small strain and low voltage), it still qualitatively captures the trend in the current. For small strain and/or low voltages, this is precisely where backward motion that was neglected in the model is most important, as well as the fact that it is where the sites (except the main binding site) are the least well-defined, see Figure~\ref{fig:CK}. We have seen from molecular dynamics simulations, as well, that there is a small, backward moving current, even at quite large voltage drops. Despite neglecting these effects, we still conclude here that Equation~\eqref{eq:model} is sufficient to understand and capture ionic transport through the graphene crown ether pore at $q_O=-0.54\,e$, as well as $q_O=-0.24\,e$ at 0.5 V and higher (with a slight modification of rates). 

{\textbf{Rate constants: }} 
The rate constants $k_b$, $k_a$, and $k_d$ depend on attempt frequencies and free-energy barriers that ions encounter during the translocation from one side of the membrane to the other~\cite{lauger1973ion, nelson2011permeation}. We will consider $k_b$ to have separate diffusion and drift components and for the other two rates to have explicit barriers. When $U_a$ and $U_d$ are the barriers to enter the pore and exit the pore, respectively, then $k_a=k_a^0\,e^{-U_a/\kt}$ and $k_d=k_d^0\,e^{-U_d/\kt}$, where $k^0_a$ and $k^0_d$ are the rate constants for barrierless transport. 

Bulk rate constant---We first consider the rate constant from bulk, $k_b=k_{b0}+\kappa_b V$, to the sites on the association side. For pure diffusion, the standard result is to take a capture radius equal to the pore radius $\a$ and solid angle $\Theta$~\cite{hille2001}, which would give
\begin{equation}
    k_{b0} \approx \Theta\,D\,c\,\a ,
\end{equation}
where $D$ and $c$ are the diffusion coefficient and the bulk ion concentration. We note that here one could argue that we should take $r_s$ (the spread of staging sites) or some modification depending on the Debye length. However, $r_s$ and $\a$ are related and, indeed, {they are equal}  in this work (as discussed above). The influence of electrostatic interactions is even less clear, as the pore rim is charge neutral on the scale of the Debye length. Thus, we consider $\a$ only, but there could be further refinement of the estimates of the model and parameters. The solid angle, $\Theta$, is generally taken to be $4\pi$ in chemical reactions~\cite{collins1949diffusion, berg1977physics} and $2\pi$ for transport through pores~\cite{behr1985carrier,wanunu2010electrostatic}.  Instead, considering the pore to be a circular disc rather than a sphere, one obtains $\Theta=4$~\cite{Hall1975, hill1975effect}. These estimates of $\Theta$ implicitly assume that the particle size is negligible compared to the capture radius. A similar estimation for sub-nanoscale pores is difficult due to the commensurate length scales involved: pore size, hydrated ion size, and Debye length are all similar. For example, L\"{a}uger pointed out that the effective capture radius of the pore can be as small as the difference between the geometric radius and the ion radius~\cite{lauger1976}, which, of course, would give rise to similar issues that we discussed in the context of the pore radius.

Assuming that a region of radius $\a$ mimics the capture of a circular disc gives $k_{b0} \approx 0.5$~ns$^{-1}$ for one molar concentration. Different assumptions about the capture geometry yield only an order one deviation in this estimate. Thus, this is in excellent agreement with that found by fitting the current data to the model, employing the pore occupancy directly from MD, which gives $0.5$ ns$^{-1}$ also. We note that the same scale of $k_b$ is used for biological ion channels~\cite{lauger1973ion}.

The driven component of the incoming rate from bulk is also inline with heuristic expectations: Ignoring diffusion and when the current is determined predominantly by the pore itself, see Equation~\eqref{eq:TotalRes}, the current will be 
\begin{equation} \label{eq:appcurr}
    I \approx \frac{V}{\gamma_p h_p/(\pi \a^2)}
\end{equation}
in the continuum limit with pore resistivity $\gamma_p$. Again ignoring diffusion, the bulk drift has to supply this same amount of current. Converting to a rate, this gives 
\begin{equation} \label{eq:kappa}
    \kappa_b V \approx \pi \a^2 V/(q \gamma_p h_p ) .
\end{equation}
Alternatively, one can think of this scenario as one where the voltage drop in bulk on one side of the membrane is $V_b\approx \pi \a \gamma_b V/(4 \gamma_p h_p)$, which comes from taking the exact---assuming a continuum with bulk resistivity $\gamma_b$---voltage drop on one side of the bulk $I\cdot R_a$ and approximating the current as in Equation~\eqref{eq:appcurr}. This partial voltage drop then supplies ions at a rate determined by its bulk, access resistance, $V_b/R_a$ (note that here $R_a = \gamma_b/(4 \a)$ as we are dealing with one side of the membrane). This yields a bulk rate identical to Equation~\eqref{eq:kappa}. The bulk resistivity is $\gamma_b=0.071\, \Omega \cdot$m for 1 mol/L KCl in rigid TIP3P water~\cite{sahu2019colloquium}. Putting in approximate values $h_p \approx 1$ nm (see, e.g., Ref.~\cite{sahu2018maxwell}) and $\a \approx 0.1$~nm yields either $1.4 \times 10^9$ ions/(V$\cdot$s) when using just the K$^+$ resistivity ($\gamma_p \approx 2 \gamma_b$) in Equation~(\ref{eq:appcurr}) or $2.8 \times 10^9$ ions/(V$\cdot$s) when using the KCl resistivity as the pore resistivity (in the bulk, we use $\gamma_b$), which is in reasonable agreement with the extracted value of $(2.2 \pm 0.2) \times 10^9$~ions/(V$\cdot$s). Again, some parameters, such as $\a$ and $h_p$, may be different, including when one is looking at different characteristics (access versus pore resistance), but at most this will give an order one change---for instance, employing $\a \approx 0.13$~nm and $\gamma_p \approx 2 \gamma_b$ would give $2.3 \times 10^9$ ions/(V$\cdot$s). In these estimates, we do allow $\gamma_p \neq \gamma_b$, but this is imposed from above rather than a consequence of free energy barriers or concentration gradients, both of which have more complex repercussions. Computing these even within a continuum picture would require a self-consistent solution, {including}  without local electroneutrality. Barriers in the pore, though, are easy to incorporate, they lower the current and thus lower the drift-induced feeding $\kappa_b$ (equivalently, they reduce $V_b$). The proximity of the estimates, though, suggests that the pore in the plateau regime is similar to that of a small open pore---``open'' meaning no free-energy features. While the pore does have energetic features for smaller strain (i.e., 2\% to 6\%, see the SM), this entails that those features are irrelevant in the sense developed above. For larger strain (8\% and 10\%), the pore is basically barrierless, even in a more strict sense (see the SM). The agreement between treating the drift rate as that in response to a small but otherwise open pore ($h_p \approx 1$ nm and $\a \approx 0.1$~nm) may be coincidental, however, as the ion crossings happen at a smaller scale in the middle of the pore ($\approx$0.02~nm). We will discuss this further below.

We note also that the values for $k_{b0}$ and $\kappa_b$ are in rough agreement with the one way rates shown in Figure~\ref{fig:rate}. Those rates seemingly would suggest a $k_{b0}$ about 4 times higher. However, these are one-way rates to cross a whole $z$-plane. Therefore, they will be larger than the rate to go into the staging sites. There are simply more fluctuations in both directions across a $z$-plane far from the pore when one is in the charge layers that maintain the potential drop. If one instead looks at the rates crossing a hemispherical surface (see the SM), the magnitudes are about a factor of two different than $k_{b0}$. This agreement is thus still only approximate, but it does suggest consistency of the model and MD data. The agreement with $\kappa_b$ is also reasonable -- the increase of the one-way rates when voltage goes from 0.1 V to 1 V is about $10^9$ ion/s to $2\times10^9$ ion/s, which agrees with the extracted $\kappa_b \approx 2.2 \times 10^9$ ion/s. 

Finally, we discuss an alternative potential interpretation (pun intended): Above, we considered rates given separately by bulk diffusion and bulk drift. However, it could be that small entrance barriers, specifically into the association-side staging sites, are giving a weakly-activated process, and hence the voltage enters through the exponent, i.e., $k_b = k_{b0} e^{\beta V/\kt}$. There are small features in the free energy around which ions would associate into the staging sites, as well as depleted ion density there, see Figures~\ref{fig:CK}~and~\ref{fig:pmf-V} (and similar figures in the SM). Using this as a fitting form also results in a reasonable fit, albeit slightly worse than the form we use above, especially at low voltage. The resulting fit parameters are $k_{b0}=(0.79 \pm 0.06)\times 10^9$~ion/s, $\beta=(1.3 \pm 0.1) \kt$/V (at room temperature), and $\tilde{k}_a=(1.0 \pm 0.4) \times 10^{10}$~ion/s, with uncertainties given by the standard error of the fit. All these numbers are inline with the heuristic estimates.

The major difference between these two interpretations is the behavior at small voltage and that the drift-based interpretation better captures the data at the smallest voltage we examine (0.1 V). Otherwise, it will be difficult to discern the exact diffusion/entrance mechanism: The expected one-sided access-induced potential drop is 1.5~$\kt$ at room temperature when the total applied voltage is 1 V (i.e., about 40~$\kt$). From a homogeneous drift theory~\cite{Hall1975}, about half of this is expected to drop within a distance $\a$ from the pore (i.e., within the ``Hille'' hemisphere~\cite{hille1970ionic})---that is, one would have to dissect small changes in the free energy and potential with voltage in the same spatial vicinity. The precision to which the calculations would have to be performed is astounding -- not just statistical precision, which can be made smaller than this, but non-scaling finite-size effects would have to be nearly completely removed~\cite{sahu2018maxwell,sahu2018golden}. Clearly, studying the temperature dependence can help further delineate these two interpretations by revealing activation energies (provided that the temperature dependence of other factors, such as the resistivity, can be accounted for), as can an even more comprehensive study including smaller voltages (where activation will be more clearly visible) and larger simulation cells (to completely---to within more than $\kt$---remove non-scaling finite-size effects~\cite{sahu2018maxwell,sahu2018golden}). While the data here favor the drift-based interpretation, it is not conclusive but it does not affect the main findings of a diffusion-limited regime around 0.1 V. {Yet another alternative model}  is to just retain $\kappa_b$ in $k_b$ (i.e., $k_{b0}=0$). This assumes that just drift is feeding ions to the pore. However, this model gives a poorer fit and is not consistent with the data. Thus, diffusive contributions from the bulk are present. 

Association rate constant---We next consider $k_a$. The model fit did not directly give us this parameter, but instead the effective association rate $\tilde{k}_a=k_a P_1^{\mathrm{eq}}=(1.2 \pm 0.3) \times 10^{10}$ ion/s. This rate is three times larger than $k_b$ at 1 V, and 30 times larger than $k_b$ at 0.1 V. Thus, the only time this component of the resistance matters is when the factor, $1-P_2$, multiplying it in Equation~\eqref{eq:model}, is small, which occurs only when both strain and voltage are small. It is difficult to estimate this parameter a priori without sufficient gymnastics as to obscure the truth of the matter. However, there are two qualitative features that support its magnitude. The first is that $P_1^{\mathrm{eq}}$ is relatively small, which can be seen from Figure~\ref{fig:CK}, probably around 1/10 or smaller. This means that $k_a$ is an order of magnitude or more larger. From Figure~\ref{fig:pmf-V}, a large value of $k_a$ is expected. There is only a small barrier around $-0.4$ to $-0.3$ nm for the $q_O=-0.54\,e$ pore, and then the ion will be driven downhill into the pore binding site. That is, we do expect a large $k_a$ for an ion already in the staging area. 

Dissociation rate constant---We next consider $k_d$. This parameter does not participate in the model fitting at all, since we instead used the computationally determined $P_2$, which enabled us to only employ the first of the three equations, Equation~\eqref{eq:P1}, in the model. However, we can employ the outgoing current from the second site, i.e., the second term in Equation~\eqref{eq:P2}, to estimate $k_d$. We can do this by noticing that $P_3$ is also small (just as $P_1$ is). Since here, $1-P_3$ is present, the estimate assuming $P_3$ is small will be less sensitive to this assumption compared to $P_1^{\mathrm{eq}}$ and $k_a$. On the plateau, this entails that $k_d P_2 = I/q = \,$constant. {Examining this relation, data point by data point, gives $k_d$ estimates between $10^9$~ion/s to $10^{11}$~ion/s, with well defined trends versus strain and voltage.} For instance, at 1 V, one obtains $k_d=2.0 \times 10^{10} e^{(0.20 \pm 0.01) s}$ ion/s, where $s$ gives the strain in percent and the confidence interval of the prefactor is [$1.9 \times 10^{10},2.2 \times 10^{10}$]~ion/s. Thus, the $k_d$ {varies from $2.0 \times 10^{10}$~ion/s at 0\% strain to $1.5 \times 10^{11}$~ion/s at 10\% strain. For completeness, the remaining voltages give fits for $k_d$ of $2.7 \times 10^9 e^{(0.28 \pm 0.03) s}$~ion/s, $8.3 \times 10^8 e^{(0.29 \pm 0.02) s}$~ion/s, and $4.3 \times 10^8 e^{(0.24 \pm 0.02) s}$~ion/s for 0.5 V, 0.25 V, and 0.1 V, respectively, with corresponding prefactor confidence intervals [$2.2 \times 10^9,3.3 \times 10^9$]~ion/s, [$7.1 \times 10^8,9.7 \times 10^8$] ion/s, and [$3.7 \times 10^8,5.0 \times 10^8$] ion/s. A reasonable fit to all the plateau data (versus voltage and strain) is $3.2 \times 10^9 V e^{(0.043 \pm 0.003) q V/\kt + (0.22 \pm 0.01) s}\,\mathrm{ions/(V\cdot s)}$ with confidence interval [$2.8 \times 10^9,3.6 \times 10^9$]~ions/(V$\cdot$s)} (note that $V$ is a prefactor out front, as well as in the exponent, since ions are driven across the pore). One takeaway from this is not only the order of magnitude of $k_d$, but that strain and voltage in this regime change barriers in the pore region by $\mathcal{O}(\kt)$. This is in agreement with estimates of how barriers change due to strain, see Equation~\eqref{eq:energy} and Ref.~\cite{sahu2019optimal}. However, it is somewhat surprising that voltages, that are 10s of $\kt$, do not change the barriers by more. The reason that this is not the case is that the pore barriers simultaneously are not playing a strong role in the resistance (i.e., they are irrelevant in the language above and thus do not self-consistently get removed by the voltage) and they occur on a scale of 0.1 nm. This means that what is relevant is the $\Delta V$ on this scale. For 1 V that evenly drops over 1 nm, this is only 4 $\kt$, i.e., only about a factor of 2 to 4 (assuming 0.2 nm over which the barrier occurs) above the actual change found in the fitted form above. We do not expect back-of-the-envelope estimates to do much better. 

To give an independent estimate of the dissociation constant for comparison, we assume that only one ion occupies the pore at a time and the translocation is driven by a constant electric field, $E_p$, across the internal pore site of length $\Delta_p$. Thus, the rate constant for exit from the pore may be estimated from the drift velocity-like picture (with an effective diffusion coefficient as an attempt frequency times an Arrhenius factor) as
\begin{equation} \label{eq:kdenvelope}
k_d =\frac{v_d}{\Delta_p}=\frac{q\,D\,e^{-U_d/\kt}  \,E_p}{\kt\,\Delta_p}  =\frac{q\,P_p\,E_p}{\kt} ,
\end{equation}
where $P_p = D\,e^{-U_d/\kt}/\Delta_p=D_p/\Delta_p$ is the permeability of the ion. Assuming that $E_p=V/h_p$ with $h_p=1$~nm (i.e., a potential drop over the effective membrane thickness that is larger than the internal pore site), we obtain $k_d^0\approx 76 \,\text{ns}^{-1}$ (for $\Delta_p=h_p$) to $190 \,\text{ns}^{-1}$ (for $\Delta_p=0.4$~nm, which is more represented of the $P_2$ site) for 1 V applied voltage and the potassium mobility $\mu_K=q D/\kt = 7.62 \times 10^{-8}$ m$^2$/(V$\cdot$s). This is in reasonable agreement with the $k_d \approx 150 \,\text{ns}^{-1}$ found above for 1 V and 10\% strain, where the latter has the smallest influence of barriers and is thus most similar to the barrier-free estimate here. Notably, however, this estimate decreases linearly with voltage. For 0.1 V, the estimate is 10 times too high compared with the one found from the MD data. However, it is clear from the $P_2$ data that occupancy is dropping faster than exponentially with voltage, meaning that $k_d$ increases faster than exponentially (note that the pore conductance, proportional to $k_d P_2$ decreases with voltage, in line with diffusion-limited expectations). The form fitted above for all plateau data assumed a form $V e^{v V}$, with $v$ as a positive constant. This form performs well and indicates that Equation~\eqref{eq:kdenvelope} is only reasonable where the voltage is not modifying the energetic landscape at all. 

Finally, we comment on the magnitude of $k_d$ compared to the bulk rate constant $k_b$. Even taking into account the effect of the potential well in the pore, the dissociation rate constant is still larger than the association rate constant ($\sim0.5\,\text{ns}^{-1}$) for most cases. For plateau data, the smallest the dissociation constant becomes a factor of two larger, but for almost all data, it is an order of magnitude larger or more. Only in the unstrained pore with $\qO=-0.54\,e$, where there is a large exit barrier, are the two rate constants comparable at a small voltage. Thus, ionic transport in the crown ether pore is---outside of the colossal mechano-conductance regime---generally controlled by the rate at which ions arrive in the pore, i.e., the diffusion and drift rates, possibly with some reflection at the pore mouth due to a small entrance barrier. The latter includes an overall barrier for pore occupation when extended to $\qO=-0.24\,e$, see Figure~\ref{fig:pmf-V}.

{\textbf{Diffusion-limited currents: }} 
The observation of diffusion-limited currents requires both that current-carrying ions spend little time in the pore and that drift component of feeding ions to the pore is small. In terms of the model, Equation~\eqref{eq:model}, we need $k_b \ll \tilde{k}_a (1-P_2)$ and that $\kappa_b V_b \ll k_{b0}$. To meet the former condition requires that  $k_b \ll \tilde{k}_a$ (i.e., entrance barriers should not be large, or otherwise $\tilde{k}_a$ will be small) and $\tilde{k}_a \ll k_d$ (to ensure that $P_2$ is not close to one), which combines to the chain of inequalities 
\begin{equation}
k_b \ll \tilde{k}_a \ll k_d .
\end{equation}
In other words, both entrance and exit barriers should be small (i.e., transport in a near barrierless regime, where ``near'' is defined in terms of how fast ions arrive at the pore from bulk and thus, under realistic conditions, even barriers in the range of 5 $\kt$ can be ``near'' barrierless for this pore, but what quantifies ``near'' depends on pore characteristics). 

When $k_b\ll \tilde{k}_a (1-P_2)$ in Equation~\eqref{eq:model}, we get $I= q\,k_b $, in which case the ionic current is fully determined by the incoming rate from bulk and is independent of pore conditions, as seen in Figure~\ref{fig:IS}. Albeit, one has to compare $k_b$ to the association rate, which not only can have a free energy barrier associated with it, but also the equilibrium occupancy of the stating site $P_1^{\mathrm{eq}}$, and thus $\tilde{k}_a$ can be quite small~itself.

The conditions above can in turn be employed to put conditions on the voltage. First consider a lower bound: Equation~\eqref{eq:kdenvelope} gives the rate at which ions cross the pore, including both the drift (due to the local electric field) and the dissociation from a pore well (if present). This rate is proportional to the voltage. Considering the chain inequality above and considering $k_d$ at $U_d = 0$ (i.e., the time spent in the pore without a well needs to be much greater than the bulk feeding---the presence of a well will only push this inequality toward not being satisfied):
\begin{equation} \label{eq:lowerbound}
k_d \gg k_b \approx k_{b0} \implies
E_p \approx V/h_p \gg \Theta\,\kt \,c\,\a\,h_p/q ,
\end{equation}
where we can take $k_b \approx k_{b0}$, since we are interested in the regime where diffusion dominates over drift. We also take $\Delta_p \approx h_p$ (this only drops an order one factor). This relation indicates that the diffusion-limited current is likely to be observed in short and narrow pores, provided that entrance and exit barriers are small. Note that, although many biological ion channels are not necessarily short compared to their width, ions can move in a single-file concerted motion via ``knock-on'' mechanisms~\cite{kopec2018direct, sahu2019optimal}, which diminishes the effective length of the channel.

An approximate upper bound on the voltage to observe diffusion limitations is, as already noted, for there to be little voltage drop in the bulk. For instance, L\"{a}uger points out that the presence of excess impermeable, or ``inert'', electrolyte increases the impact of diffusion limitations~\cite{lauger1976}, a fact that occurs in our pore (i.e., Cl$^-$ is inert). This is due to the fact that an impermeable electrolyte shifts the balance of pore and bulk resistance, making the former much larger relative to the latter. Assuming a cylindrical pore, the pore resistance is dominant if $h_p/\pi \a \gg \gamma_b/2\gamma_p$, where $\gamma_b$ ($\gamma_p$) is the resistivity in bulk (pore). For larger graphene pores, and even some nanoscale ones, this condition is unlikely to be true. In fact, access resistance is larger than the pore resistance for most sizes of graphene pores, and thus the majority of the voltage will drop in the bulk. In such a case, the current will be limited, not by diffusion but mostly by~drift.

In the graphene crown ether pore, however, the effective pore radius is around 0.1 nm and, when under strain, near barrierless in the sense used above (there may be barriers and wells, but the prefactors---the transition rates or attempt frequencies---are still determining the hierarchy of rate scales). Using the same estimate to find $\kappa_b$ as above, where we assume a homogeneous, continuum medium both inside and outside the pore, with the pore resistance the dominant factor, we obtain
\begin{equation} \label{eq:upperbound}
    \kappa_b V \approx \pi \a^2 V/(q \gamma_p h_p) = \pi \a^2 V c_p \mu_p/h_p \ll k_{b0} \approx \Theta\,D\,c\,\a .
\end{equation}
This upper bound can be derived directly from the steady state Nernst--Planck equation, assuming hemispherical symmetry (i.e., with only a radial component) and a homogeneous medium. There, one~wants
\begin{equation} \label{eq:NPcond}
    \frac{\partial c}{\partial r} \gg \frac{q c}{\kt} \frac{\partial \Phi}{\partial r},
\end{equation}
to have the diffusion contribution much larger than the drift, where $\Phi$ is the electric potential. Taking the pore mouth to be a hemisphere with radius $a_p$. The RHS is $q c a_p V_b/(r^2 \kt)$. The LHS is $\Delta c a_p/r^2$, with $\Delta c$ the concentration bias between the bulk (infinitely far from the pore) and the hemispherical pore mouth. This presumes that the diffusive and drift components are decoupled.

Assuming further that the staging site has zero occupancy (and thus zero concentration), this gives the maximum diffusion contribution and Equation~\eqref{eq:NPcond} results in 
\begin{equation} \label{eq:Vbbound}
    q V_b \ll \kt.
\end{equation}
This relation is interesting in itself. Its simplicity is due to Einstein's relation of mobility and diffusion coefficients, which results in additional factors dropping out, and due to comparing a maximum diffusive current occurring at the largest concentration gradient with the maximum drift current occurring at zero concentration gradient. Equation~\eqref{eq:Vbbound} indicates that for drift to be negligible, the voltage drop in bulk has to be less than the thermal energy. The latter ``drives'' the diffusion. It should be larger than the drive of the drift current from bulk to the pore. Plugging in the form of $V_b$ assuming a bulk potential drop in the presence of a dominant pore resistance (see just below Equation~\eqref{eq:kappa}) gives the same inequality as Equation~\eqref{eq:upperbound} up to order one factors.

Rewriting Equation~\eqref{eq:upperbound} together with Equation~\eqref{eq:lowerbound}, assuming $c_p=c$ and $\mu_p=\mu$ (i.e., that these two quantities are equal to their bulk), to obtain a two-sided inequality for $V$ yields
\begin{equation} \label{eq:difflimits}
    \Theta\,\kt \,c\,\a\,h_p^2 \ll q V \ll \Theta\,\kt\,h_p /(\pi \a) .
\end{equation}
This foundational relation gives one of the main predictions of this paper: Diffusion-limited currents appear within a sweet spot when free energy features are irrelevant. For the graphene crown ether pore, the voltage should be between about 6 mV and 300 mV. At voltages higher than this range, drift will be important and, below this range, ions will not be removed from the pore region fast enough to create a concentration gradient (and free energy features will also become relevant). The simulations and modeling validate the upper bound (at 0.25 V, the bulk drift and diffusion contributions are roughly equal), but they do not address the lower bound (in any case, free energy barriers will likely be relevant at 6 mV {for the $q_O=-0.54\,e$ pore}, as they are with the $q_O=-0.24\,e$ pore still at 100 mV and 250 mV, the relevance of which is inconsistent with the assumptions leading to  Equation~\eqref{eq:difflimits}). This range includes the {voltage, 0.1~V}, that we see the strongest diffusion limitations, whereas at higher voltages, drift starts to determine the current. The pore charge is important, as it determines when free energy features are irrelevant (e.g., {at 0.1~V but} 0\% strain, the free energy landscape is dominant). Around 0.1 V is a typical value for graphene pore experiments, {small}  enough to not degrade the membrane (i.e., 0.5 V and higher will start to see membrane degradation), but large enough that typical currents are in the 10 s of picoampere or more (while the time-resolution is irrelevant to measuring the dc conductance, we do note that pin hole leaks or other factors can set a baseline resolution of the current, around 0.5 pA at 0.1 V (see Ref.~\cite{garaj2010} where such currents in ``as-grown'' membranes could vary by an order of magnitude from membrane to membrane). We note that the rate model that we are developing cannot be used at very small voltages, as it includes only one way currents in the pore which cannot capture the approach to equilibrium as $V\to0$.

In any case, there should be a very small drift current in the bulk when the applied voltage is around 0.1 V and it should start to become comparable to the diffusive component at about 0.25 V and dominant for higher voltage. We can make this quantitative using the fit to the model in Equation~\eqref{eq:model}. For instance, at 0.1 V, the unstrained membrane has an effective pore associate rate, $\tilde{k}_a (1-P_2)$ of 12/\SI{}{\micro\second} due to the presence of a localized ion that creates a many-body blocking effect (i.e., $P_2 \approx 1$). This effective rate increases to 11/ns for 10\% strain. Meanwhile, the diffusive rate is 0.5/ns and the drift rate is 0.2/ns. Thus, diffusion supplies ions over drift by more than a factor of 2 over the whole range of strains, and already at 2\%, the strain is smaller (though comparable) to the pore association rate (about 1.9/ns).  At 0.25 V, the diffusive and drift components are comparable at 0.5/ns and 0.55/ns, respectively. These values are slightly more than the 0.25/ns effective pore association rate at 0\% strain, but are the controlling factors for essentially all strains at 2\% and above. 

Since the smallest voltage we consider, 0.1 V, has smallest bulk drift contribution, we can postulate that the plateau resistance is the closest to $\gamma_p h_p/(\pi \a^2)$ (i.e., without any access component). Employing $\gamma_p/2=\gamma_b=0.071 \Omega\cdot$m ($\gamma_b$ is the resistivity of 1 mol/L KCl in TIP3P water, see Ref.~\cite{sahu2019colloquium}) and $h_p \approx 1$ nm, this gives $\a \approx 0.11$~nm, in agreement with the effective pore radius. This is unexpected, since the potassium ions cannot make use of the full pore area for transport and there are diffusion limitations. There may be several factors that conspire to give this agreement. One is that the pore rim is not fixed but can instead move, so that $\a$ can be bigger than a priori expectations. This does not, however, seem to be the case, since the density plots show that ions are translocating closer to the origin than 0.1 nm. Another factor is the role of $h_p$. The effective thickness may be smaller than 1 nm (its value for unfunctionalized graphene pores~\cite{sahu2018maxwell}). Moreover, while transport veers toward barrierless transport, the pores are not becoming barrierless in the strict sense for either $q_O=-0.54\,e$ (until high strain, see the SM) or $q_O=-0.24\,e$ (see Ref.~\cite{sahu2019optimal} for the discussion of the latter case). However, localized binding sites can give a rate that is similar in  magnitude to free diffusion through the pore constriction, or even a higher rate, because, while ions have to jump out of the well and the barrier height thus suppresses the rate, there is still a large prefactor, since the ion is fluctuating rapidly. The enhanced density can push the currents higher than expected based on just an open area. Whether we should think about the pore as an open pore of radius 0.1 nm or whether it is a pore of radius 0.02 nm with an enhanced density due to binding, is an interesting question. Evidence---specifically the higher concentration in a smaller spatial region---suggests the latter. However, we only point out that there is still broad agreement between these two perspectives and they only inform us how we should dissect the pore resistance $R_p$ into component pieces (meaning, the utility of the perspectives is limited). 

It is to be noted that the access resistance in an MD simulation (or any other method) depends on the simulation cell size: one can make it arbitrarily small (using a wide and short cell) or large (using tall and narrow cell)~\cite{sahu2018maxwell}. Therefore, in order to match experimental conditions, which effectively has an infinite bulk, one has to {exert} great care. In our simulation, we chose the simulation cell aspect ratio to be the golden aspect ratio~\cite{sahu2018golden}, which ensures that the access resistance represents the infinite, balanced bulk resistance. Without taking this approach, one could not examine the bulk diffusion and access limitations. {In the SM, we show results of simulations for several different voltages and strains, showing that the golden aspect ratio gives converged currents, ones where the bulk is properly included}.

\section{Conclusions}
Diffusion-limited ionic currents are commonly observed in biological channels because they can provide the necessary conditions: a large pore resistance (compared to access resistance) due to the small pore radius, but also a high permeability of ions due to the functional groups that facilitate the transport of ions~\cite{Doyle98-1}. However, diffusion limitations have not been studied systematically in synthetic nanopores, since it is difficult to replicate the permeability of biological ion channels. In this regard, strained synthetic pores may provide a platform, not only to investigate the competition of dehydration and electrostatic interactions within precision atomic constructions that lead to optimal transport characteristics~\cite{sahu2019optimal}, but to investigate diffusion and entrance effects in ionic transport. 

We have shown that there is broad agreement between a simple, many-body model developed here for the $q_O=-0.54\,e$ pore and the all-atom simulations, encompassing not only the residuals and current fit, but also with the fit parameters themselves and independent estimates. This agreement suggests that, with strain, this pore transitions from a barrier-limited pore with current dictated by many-body mechanisms (i.e., a well with a localized ion that blocks the pore), to one equivalent to an open tiny pore. The pore still has a free energy structure, but this structure is irrelevant in the plateau regime: At small voltages, current-limiting regions of the landscape---dictated both by the barrier scale and the kinetics---will appear. Larger voltages will start to self-consistently remove those limiting regions by the counteracting local voltage drop. Other regions of the landscape will start to be limiting, and those regions will subsequently be washed out. For a given strain and voltage, though, the type of behavior observed depends on the pore charge and ion dehydration energy. In the particular pore here, the extent of the bulk-limited region reflects whether the unstrained pore has an internal (dehydration-dominated) barrier or (electrostatically stabilized) well.  

Therefore, the transition from the barrier-limited to the diffusion-limited regime gives the opportunity to experimentally delineate and constrain the electromechanical environment of the pore, thus pushing further the limits of employing synthetic pores to understand complex mechanisms in sub-nanoscale ion transport. Functionalized pores in two dimensional membranes are thus simultaneously complicated enough to display a wide-range of ionic phenomena seen in biological pores and simple enough to be amenable to direct modeling. Moreover, if other information can be experimentally determined, such as the (equilibrium) pore occupancy ($P_2$ here), then measurement will enable the extraction of kinetic rates and barriers via modeling. In other words, the graphene crown ether pore is about as simple a sub-nanoscale pore as possible. Yet, it displays a wide variety of behavior: single versus many-body ion competition, optimality, diffusion limitations, relevant versus irrelevant features, etc. Its behavior, for instance, will enable quantifying aspects of transport, such as the role of precision atomic placement and charge in biological systems, and a theoretical understanding of what ``near barrierless'' entails in particular pores. This area is vast, and pores in 2D membranes will provide the landscape for a systematic experimental exploration and validation of theoretical models of sub-nanoscale pores and biological channels.

\vspace{6pt}
\authorcontributions{S.S. performed the numerical calculations. Both authors modeled and analyzed data, wrote the manuscript, and clarified the ideas. All authors have read and agreed to the published version of the manuscript.}

\funding{{This research received no external funding}.}

\acknowledgments{
The authors thank D. Hoogerheide, J. A. Liddle, J. Majikes, and J. Elenewski for comments on the manuscript. S.S. acknowledges support under the Cooperative Research Agreement between the University of Maryland and the National Institute of Standards and Technology Physical Measurement Laboratory, Award 70NANB14H209, through the University of Maryland. }

\conflictsofinterest{The authors declare no conflict of interest.} 

\abbreviations{The following abbreviations are used in this manuscript:\\

\noindent 
\begin{tabular}{@{}ll}
MD & Molecular Dynamics\\
ABF & Adaptive Biasing Method\\
SE & Standard Error
\end{tabular}}

\reftitle{References}


\end{document}


\global\long\def\mr#1{\mathrm{#1}}
\global\long\def\qO{q_\mathrm{O}}
\global\long\def\prg#1{\noindent{\bf #1}}
\global\long\def\kt{k_B T}
\global\long\def\Va{V_\text{ext}}

\renewcommand{\thesection}{S\arabic{section}}
\renewcommand{\thefigure}{S\arabic{figure}}
\renewcommand{\thetable}{S\arabic{table}}
\renewcommand{\theequation}{S\arabic{equation}}

Here, we give additional results, including figures that show the behavior across the full range of parameter space (e.g., versus voltage and strain) that we study. 

\begin{figure}[H]
\centering
\includegraphics{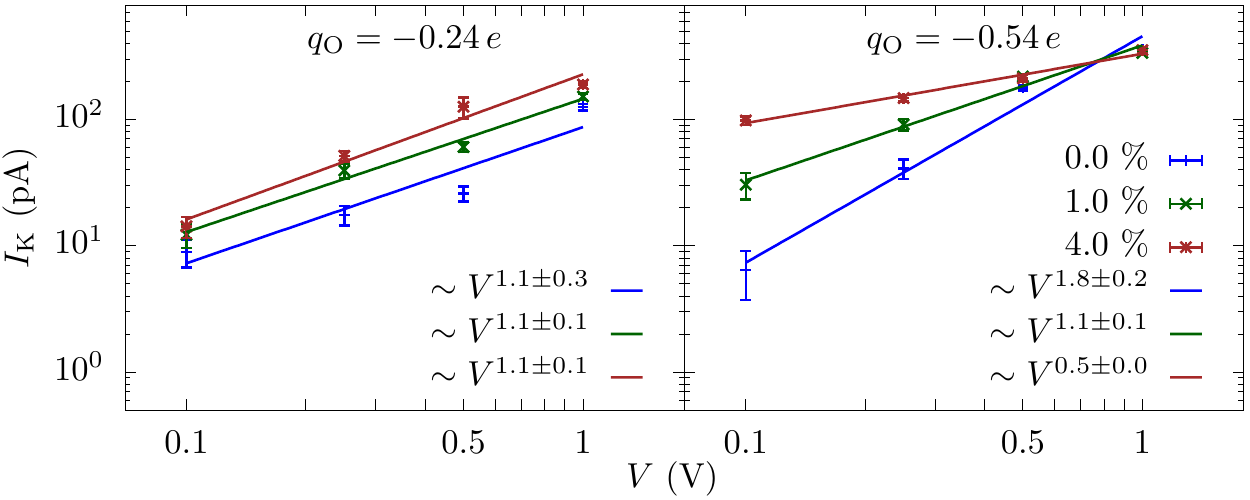}
\caption{{\bf IV-characteristics.} K$^+$ current through graphene crown ether pore at different strains in 1~mol/L  KCl versus voltage. When current is limited by barriers, the IV-characteristics are super-linear and, when current is limited by diffusion, the IV-characteristics are sub-linear. Linear behavior may appear due to the cancellation of the two effects. While some of the data fit well to a simple power law, there are prominent features due to the rather complex dependence of energetics on voltage. The error bars are plus/minus one SE from five parallel runs.  \label{fig:IV} }
\end{figure}

\prg{IV characteristics:}
We now look at the consequence of diffusion limitations and translocation barriers in the IV characteristics of the pore. In Figure~\ref{fig:IV}, we plot the IV characteristics along with fits of the form $I=a\,V^b$, where $a$ and $b$ are positive constants. For $\qO=-0.24\,e$, the best fit gives an Ohmic to slightly superlinear relation. This is expected when the current is mostly limited by large translocation barriers. The fit is poor, however, particularly for the 0~\% strain case. This is a consequence of the highly competitive nature of transport in this case, where dehydration and electrostatic interactions are giving a fairly complicated free energy landscape, on top of which voltage eventually---at $\Va=1$~V---washes out the free energy features and the current becomes much larger than an Ohmic relation would predict.

For the $\qO=-0.54\,e$ pore, we see three different regimes of IV characteristics depending on the strain. For small strain, the current is super-linear, i.e., $b>1$; for large strain, the current is sub-linear, i.e. $b<1$; and, at some intermediate strain it is nearly linear $b\approx 1$.  If the current is fully diffusion-limited, then we should get $b=0$. However, the current we observe is not fully diffusion-limited, i.e., it does increase with voltage, but sub-ohmically (see Figure~\ref{fig:IV}). Nonetheless, the weak dependence of current on the voltage at larger strain, with the channel tending to empty out as voltage increases, implies that the current has diffusion limitations.  This kind of saturation, where the channel empties out as voltage increases, is similar to that in the MaxiK  channel~\cite{nelson2011permeation}.

The diffusion limitations and translocation barriers have opposite effects in the IV characteristics. The diffusion limitations will make the IV characteristics sublinear (less dependent on voltage). Translocation barriers, on the other hand, will make the current superlinear because the applied field helps overcome the potential barrier. For small voltages, IV characteristics in the barrier limited regime can be linear, but for the large range (0.1 V- 1.0 V) we investigate here, it should be superlinear. For small strain, translocation barriers are the dominant factors and thus we see the superlinear behavior. As we increase the strain, the translocation through the pore becomes barrierless for $\qO=-0.54\,e$. Thus, the ionic current becomes sublinear (diffusion-limited). The approximately Ohmic behavior we observe for 1~\% strain looks to be a fortuitous cancellation of the two effects.

\begin{figure}[H]
    \centering
    \includegraphics{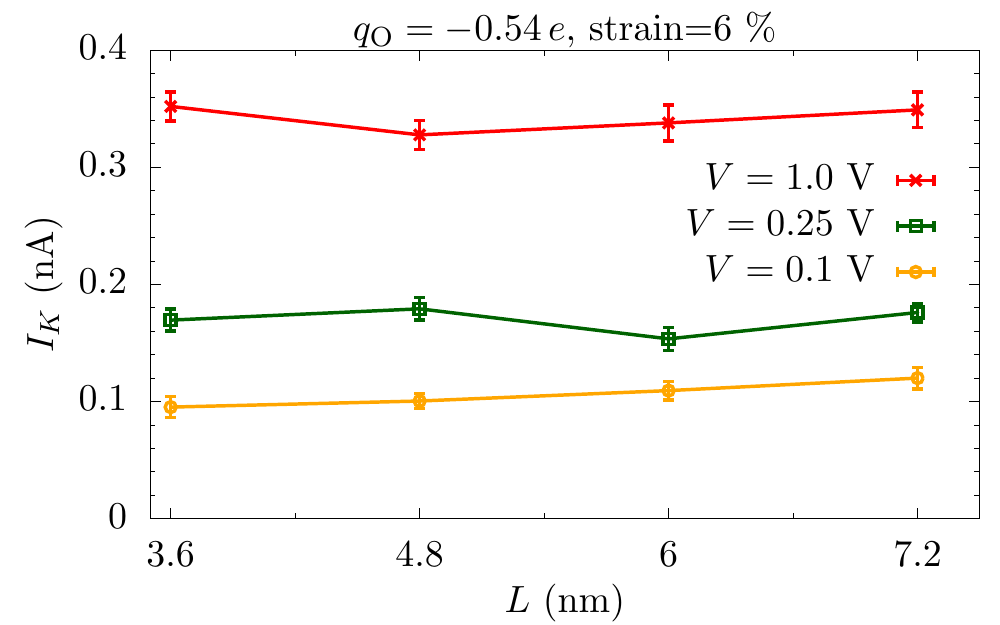}
    \caption{Potassium current through the crown ether graphene pore with various simulation cell cross-sectional lengths but with the same aspect ratio ($H/L\approx1.2$) at $\qO=-0.54\,e$ and strain 6~\%, showing that the ionic current does not vary with simulation cell size as long as the aspect ratio is kept at the golden aspect ratio. The error bars are plus/minus one SE from five parallel runs.}
    \label{fig:IL}
\end{figure}

\prg{Golden aspect ratio: }
The role of the bulk in determining resistance---specifically in access or diffusion limitations---requires a careful treatment of the simulation cell, as the bulk only slowly converges. We thus employ the golden aspect ratio method~\cite{sahu2018maxwell,sahu2018golden}. This employs a special aspect ratio that, after the disappearance of non-scaling finite-size effects, converges immediately to the infinite bulk limit. Figure~\ref{fig:IL} shows the current versus the simulation cell cross-sectional length (with height proportional to this length according to the golden aspect ratio). While there is some variation in the current over the length scales shown, it is mostly within the statistical error bars (from five parallel runs). Some of this variation may be non-scaling finite size effects, but overall the change is within expected errors.

\newpage

\begin{figure}[h]
\includegraphics{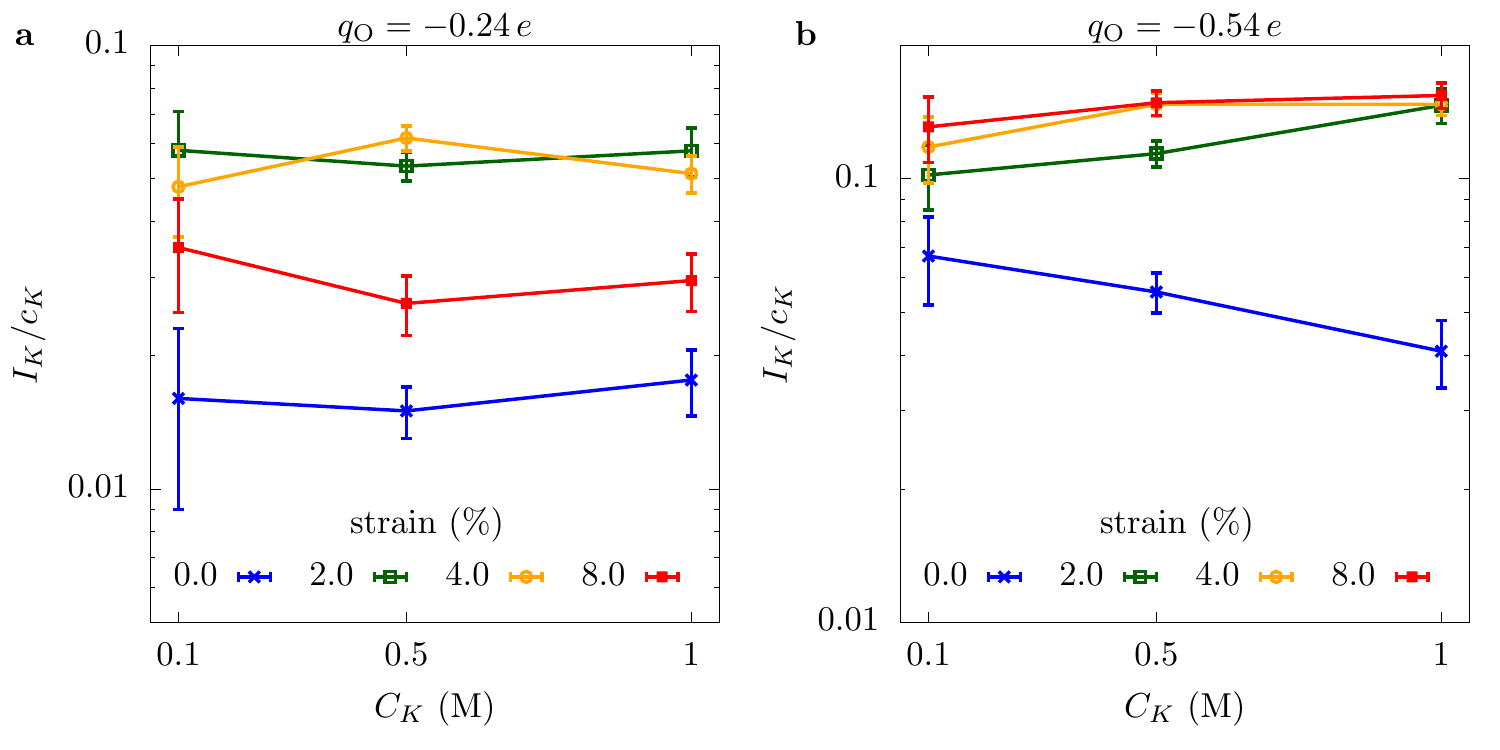}
\caption{Potassium current normalized by the bulk ion concentration versus the concentration. $I_\mathrm{K}/C_\mathrm{K}$ remains constant for all the cases where the ionic current is limited by association rate. Only for the unstrained pore with $\qO=-0.54\,e$, where ionic current is influenced by the dissociation rate, does the normalized current decrease with concentration. The error bars are plus/minus one SE from five parallel runs. \label{fig:IC}}
\end{figure}

\prg{Concentration effects: }
In the main text, we discuss how for $\qO=-0.54\,e$ at small strain and low voltage the current displays many-body blockade effects. Otherwise, the other parameter regimes are single-ion transport. One way to see many-body effects is to study the concentration dependence of the current. Figure~\ref{fig:IC} shows the normalized current versus the concentration. For $\qO=-0.54\,e$ at small strain, this normalized current decreases versus concentration, which suggests that many-body effects are at play (this dependence is just outside the error bars). Increasing the concentration does not increase the current proportionally since the main pore site is already occupied and is preventing further current flow. All other cases show only small variations or small upward trend, albeit barely out of the range of the statistical error bars from the five parallel runs.

\newpage

\begin{figure}[h]
\centering
\includegraphics{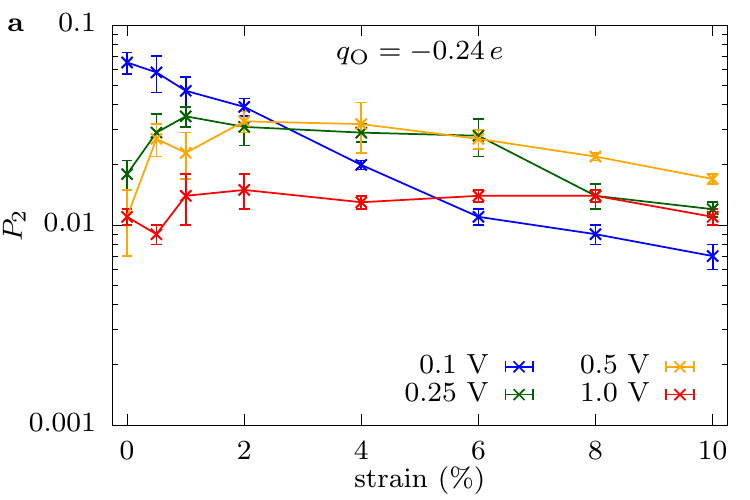}
\includegraphics{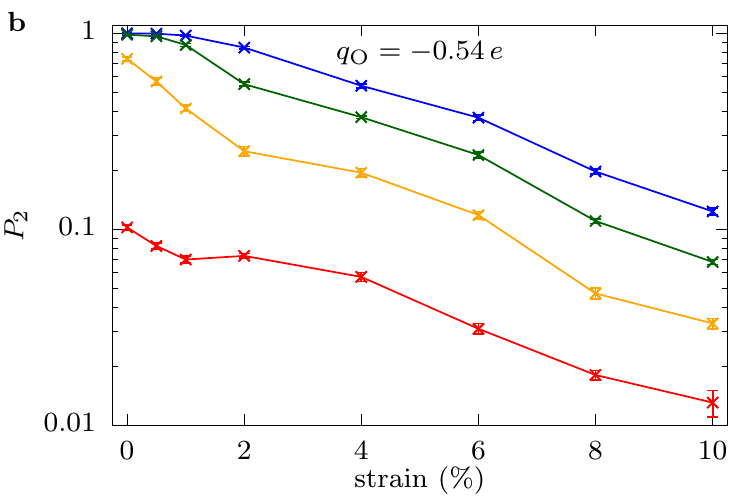}
\caption{Average number of K$^+$ ion in the pore ($|z|\leq 0.2$ nm) versus strain for {\bf a.} $\qO=-0.24\,e$ and {\bf b.} $\qO=-0.54\,e$ and for various applied bias. Note that this data includes 0.5 \% and 1 \% strain, where the main text only includes data at 2 \% increments. \label{fig:NK}}
\end{figure}

\prg{Occupancy: }
Within the main text, we develop a three site model for the current through the $\qO=-0.54\,e$ pore. The form of this model is motivated by concentration and rate data. However, it still has several parameters. To reduce the number of parameters, we compute the occupancy of the main pore site versus strain and voltage. As an input, this allows a more rigorous treatment of the fitting of the MD current data to the model. Figure~\ref{fig:NK} shows the potassium occupancy of the main pore site versus strain for the four voltages we study. The occupancy in the $\qO=-0.24\,e$ pore does not (except at low voltage) display any well-defined trends. At low voltage and within the main pore region, the free energy is increasing (the satellite barriers are decreasing), giving rise to an exponential decrease in pore occupancy. For other voltages, there is a more complex interplay of strain and voltage within the free energy landscape.  However, for $\qO=-0.54\,e$, there is a clear exponential decrease in the pore occupancy with strain due to the raising of the bottom of the potential well. Moreover, the occupancy decreases a bit faster than exponential with voltage (in the main text, we model the dissociation time as $V e^{v V}$, with $v$ some positive constant, which is both physically motivated---the exponential represents a decrease in barrier height and the pre-exponential the local field driving the ions---and works well). 

\begin{figure}[H]
\centering
\includegraphics{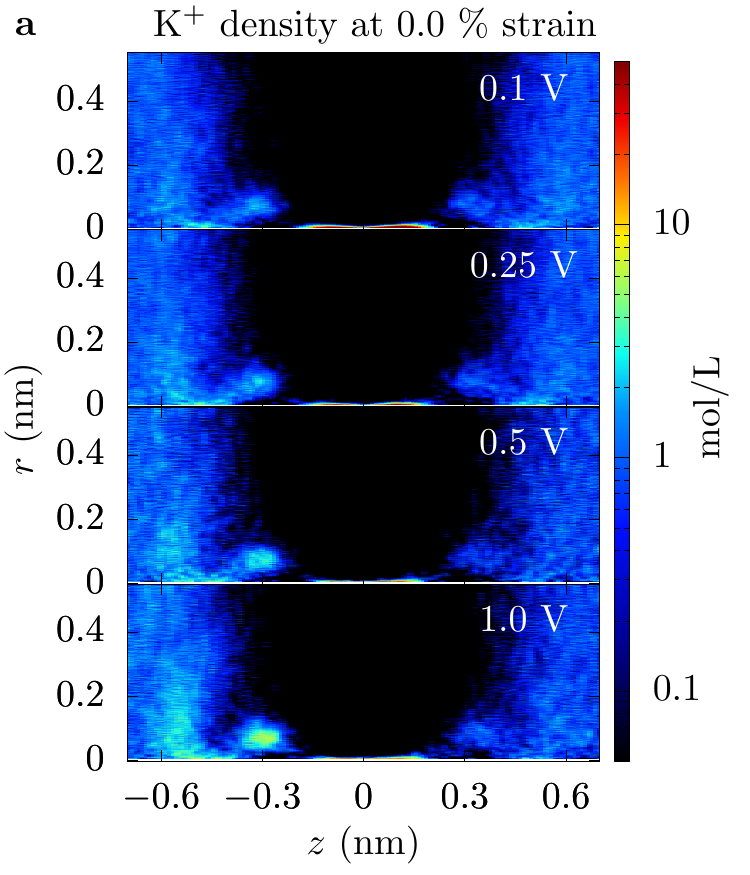}
\includegraphics{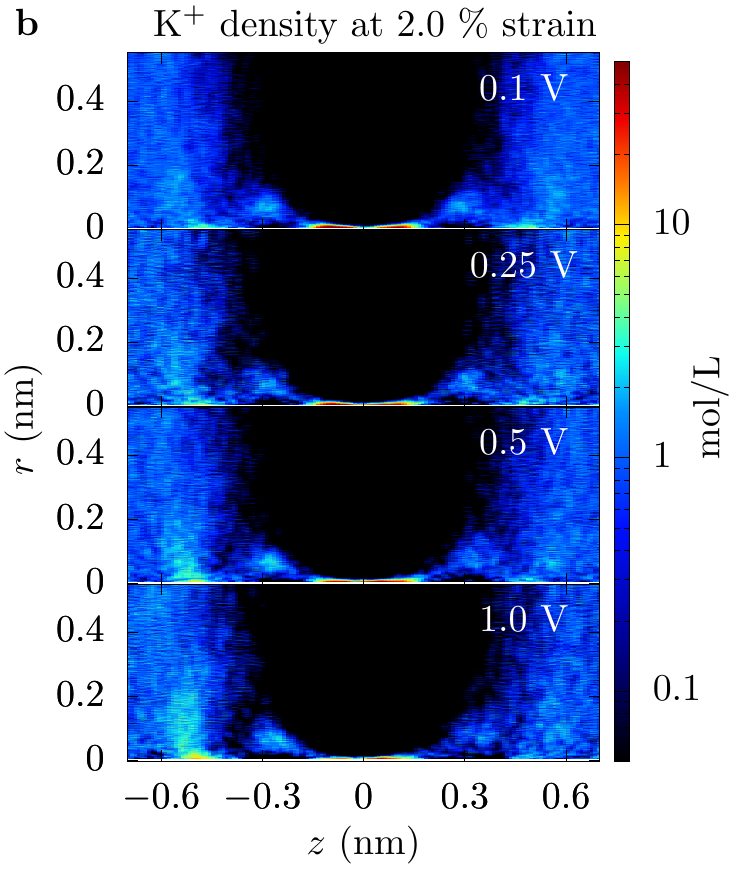}
\includegraphics{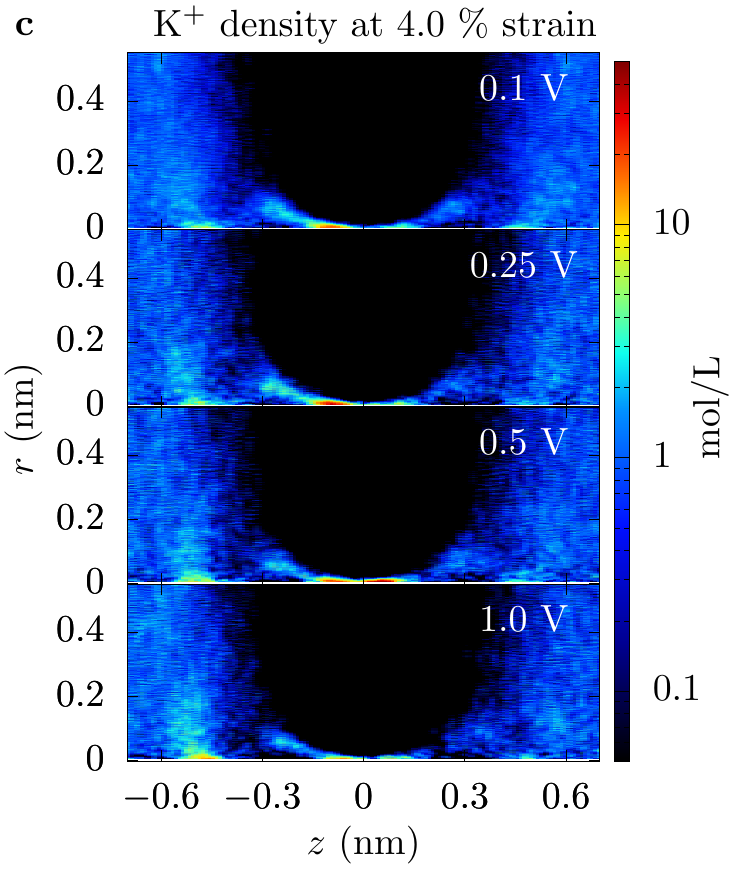}
\includegraphics{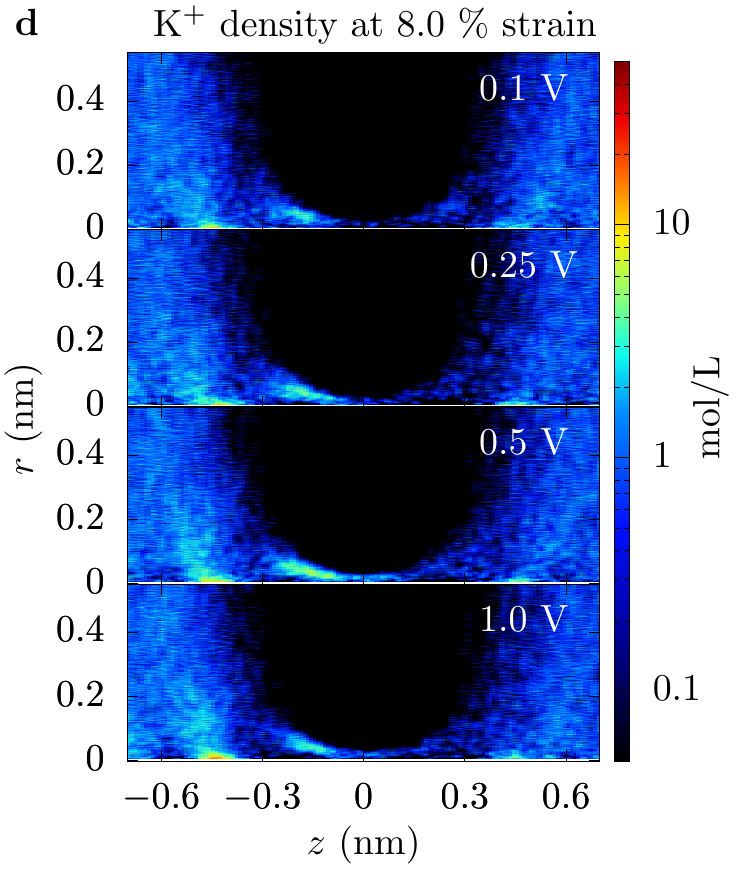}
\caption{Concentration of potassium ions near a graphene crown ether pore with $\qO=-0.24\,e$ at {\bf a} 0~\%, {\bf b} 2~\%, {\bf c} 4~\%, and {\bf d} 8~\% strain for various voltages.  \label{fig:CKrzq024}  }
\end{figure}

\begin{figure}[H]
\centering
\includegraphics{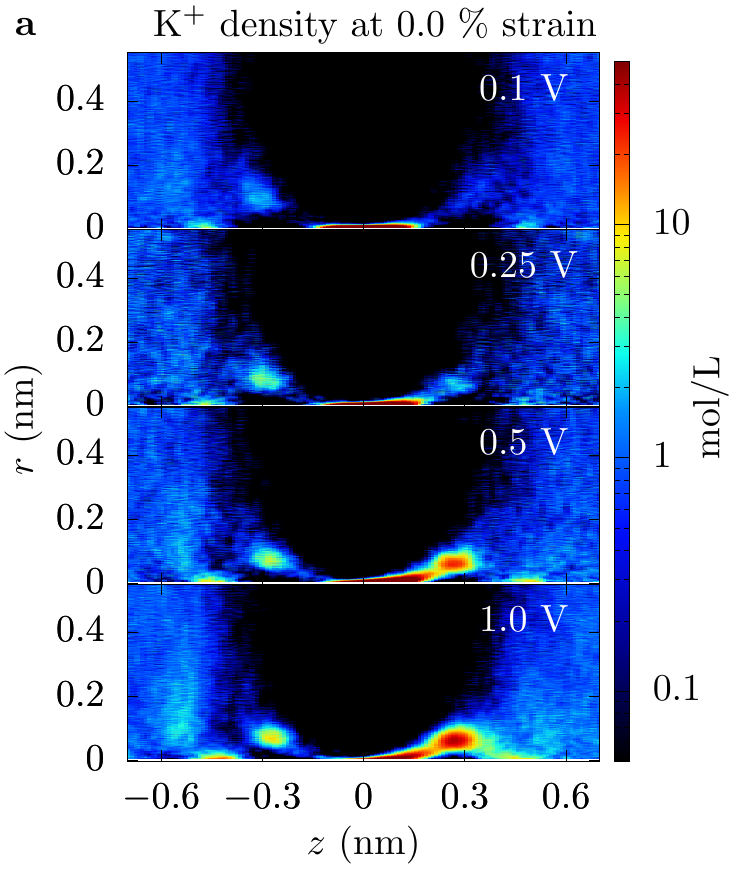}
\includegraphics{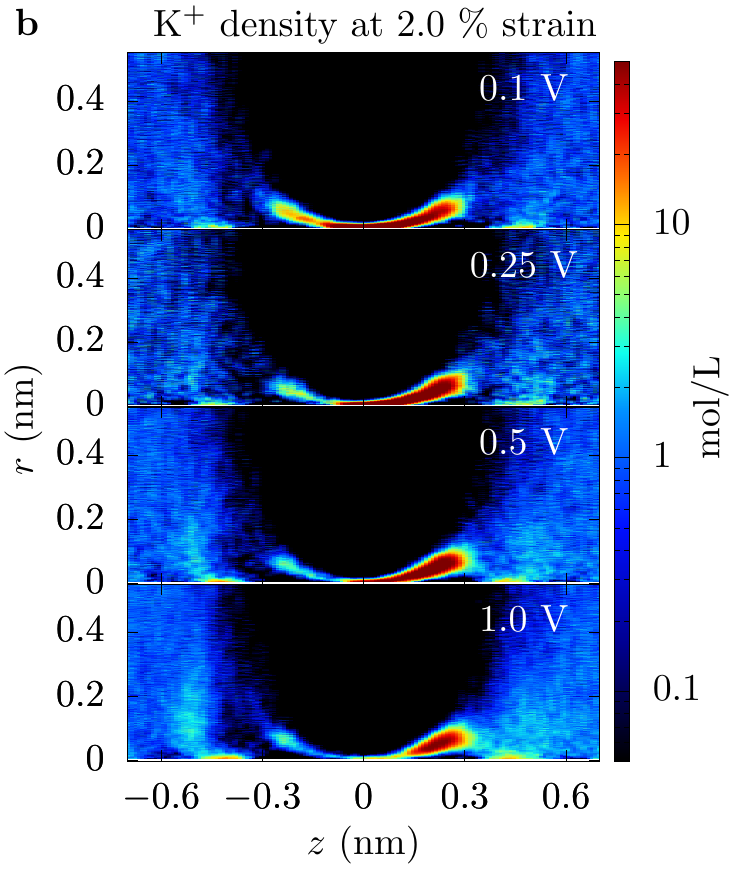}
\includegraphics{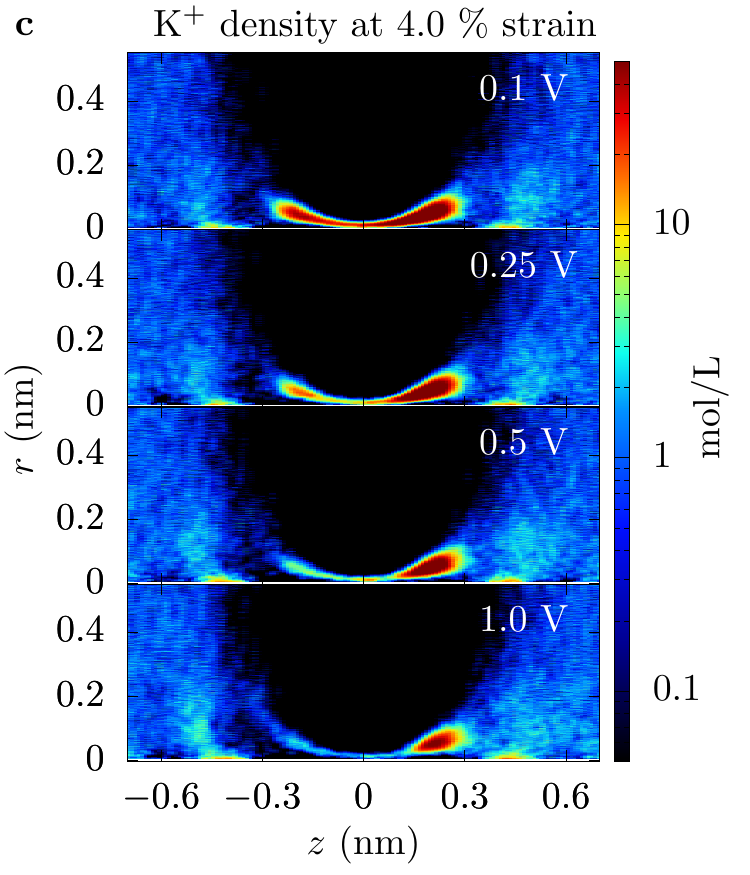}
\includegraphics{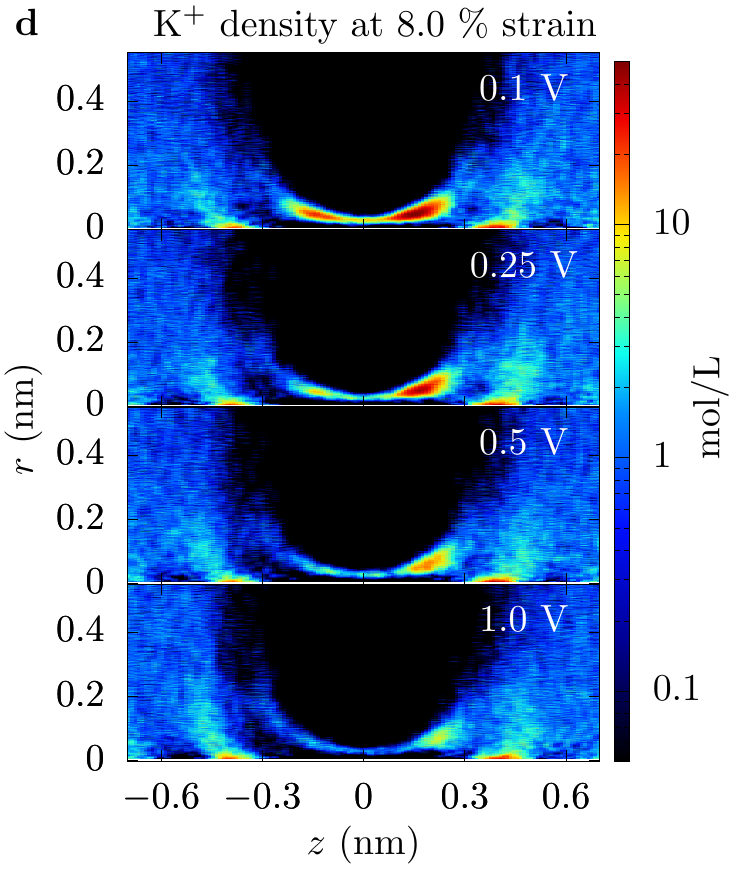}
\caption{Concentration of potassium ions near a graphene crown ether pore with $\qO=-0.54\,e$ at {\bf a} 0~\%, {\bf b} 2~\%, {\bf c} 4~\%, and {\bf d} 8~\% strain for various voltages.  \label{fig:CKrzq054} }
\end{figure}

Here, we also show additional concentration data versus spatial position, see Figure~\ref{fig:CKrzq024} and Figure~\ref{fig:CKrzq054}. This shows where the concentration can increase or decrease with voltage depending on the pore characteristics. It also motivates the three site model we take in the main text.

\begin{figure}[h]
    \centering
    \includegraphics{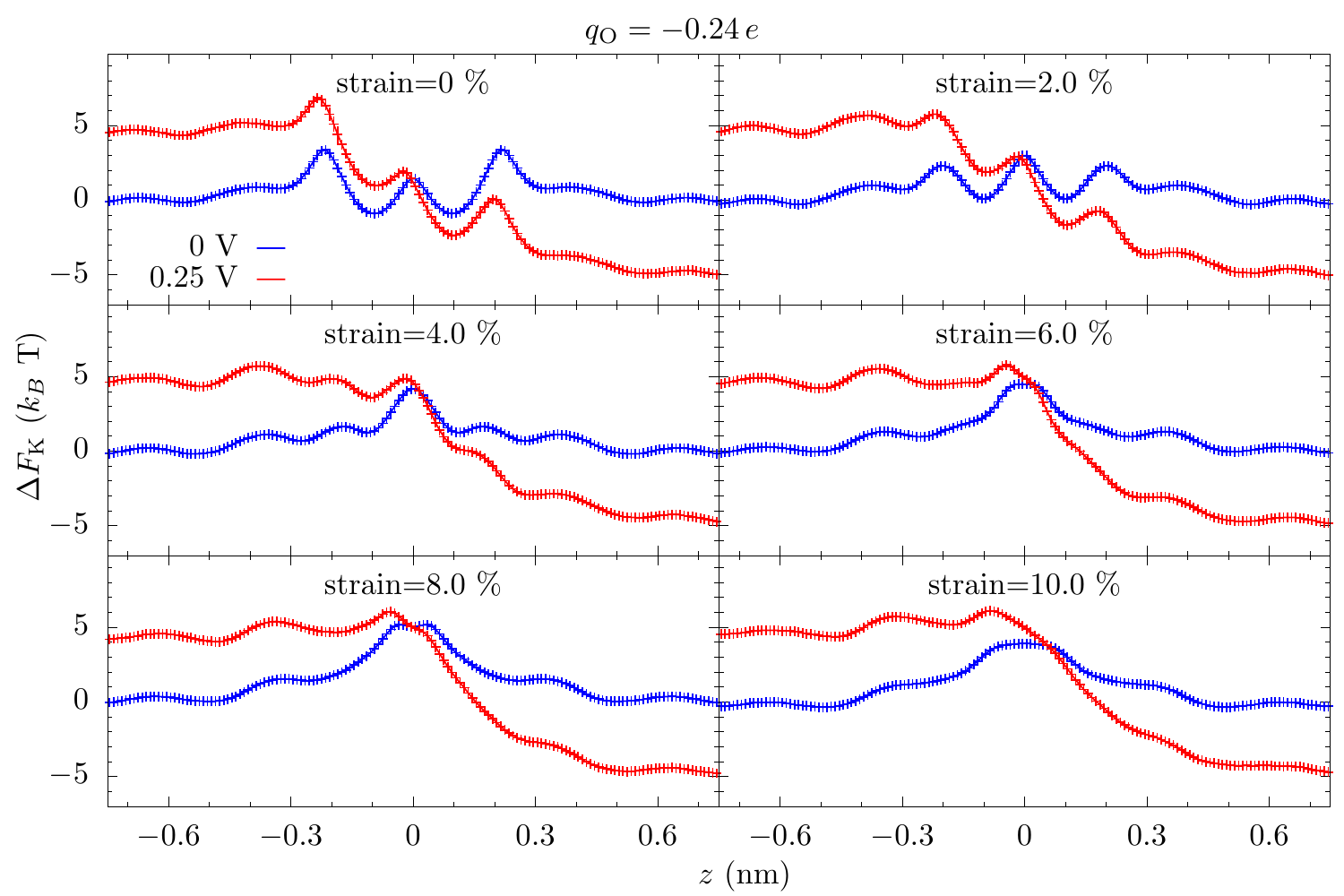}
    \caption{The free-energy profile of K$^+$ going through a graphene crown ether pore with $q_0=-0.24\,e$ at various strains for equilibrium and non-equilibrium ($\Va=0.25$~V) cases. The error bars are plus/minus one SE from five parallel runs. 
    \label{fig:pmf024}}
\end{figure}

\begin{figure}[h]
    \centering
    \includegraphics{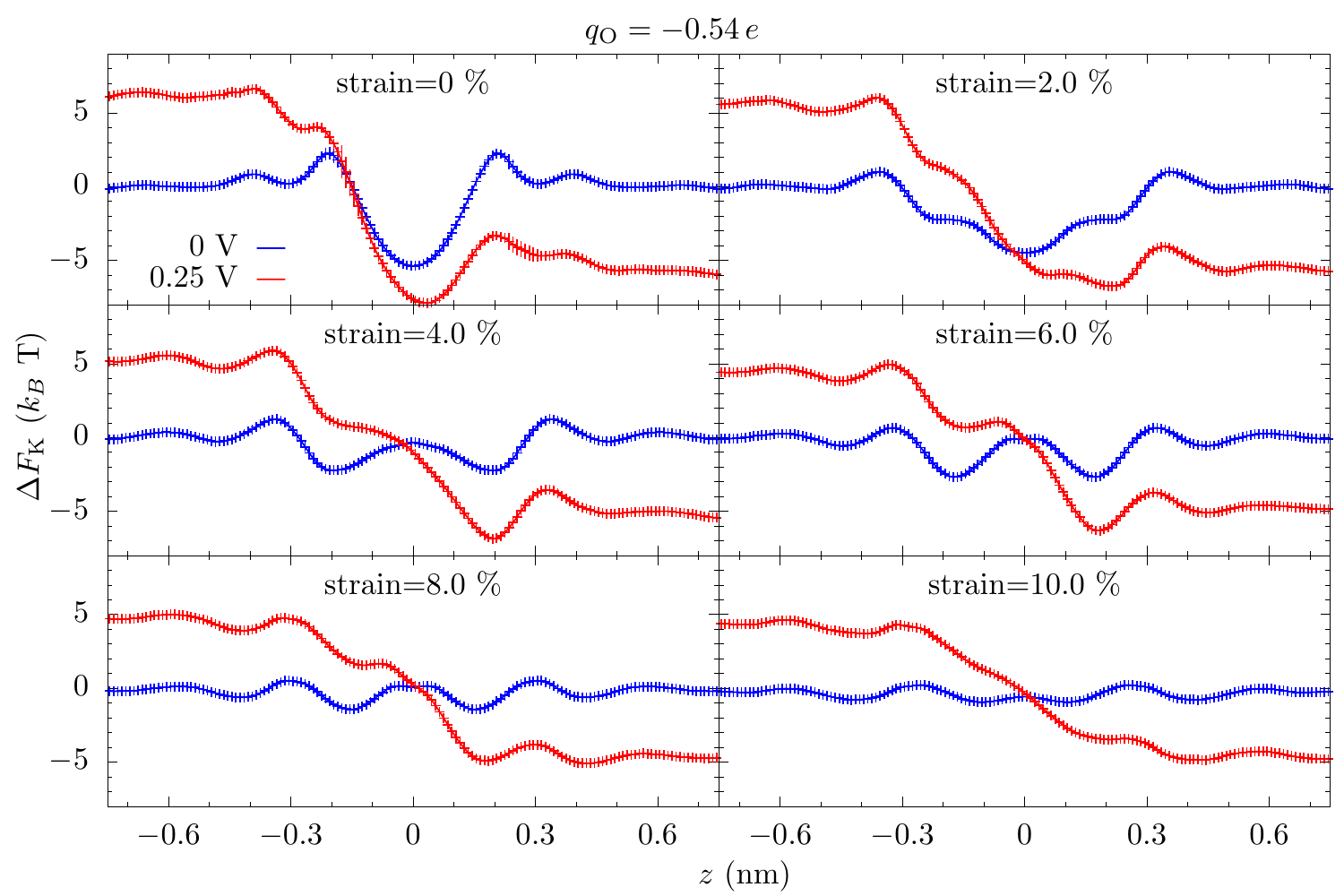}
    \caption{The free-energy profile of K$^+$ going through a graphene crown ether pore with $q_0=-0.54\,e$ at various strains for equilibrium and non-equilibrium ($\Va=0.25$~V) cases. These plots show that the $q_0=-0.54\,e$ pore veers toward a barrierless configuration, with 10 \% strain removing barriers nearly completely. Moreover, it also demonstrates that at 0.25 V, most of the cases (all except 0 \% strain) have irrelevant features in the free energy landscape, ones that remain unchanged when voltage is brought to 0.25 V from 0 V. The error bars are plus/minus one SE from five parallel runs. 
    \label{fig:pmf054}}
\end{figure}

\newpage
\prg{Free energies and electrostatic potentials: }
Figures~\ref{fig:pmf024} and ~\ref{fig:pmf054}  show additional free energy profiles, as well as a larger range of $z$. These demonstrate that there are indeed irrelevant features in free energy. Specifically, $\qO=-0.54\,e$ has a feature at 0.2~nm for most strains that changes little when the voltage is taken from 0 V to 0.25 V. Moreover, the free energy is near barrierless for the largest strain examined. For $\qO=-0.24\,e$, the barrier in the middle of the pore increases with strain until somewhere between 4~\% and 6~\% strain, after which the peak mostly broadens and then decreases. This is due to the decrease of the electrostatic interaction initially being unable to compensate for dehydration. At 0.25 V, the pore is nearly barrierless at high strain, as the bias effectively wipes out the main feature.

\begin{figure}[h]
    \centering
    \includegraphics{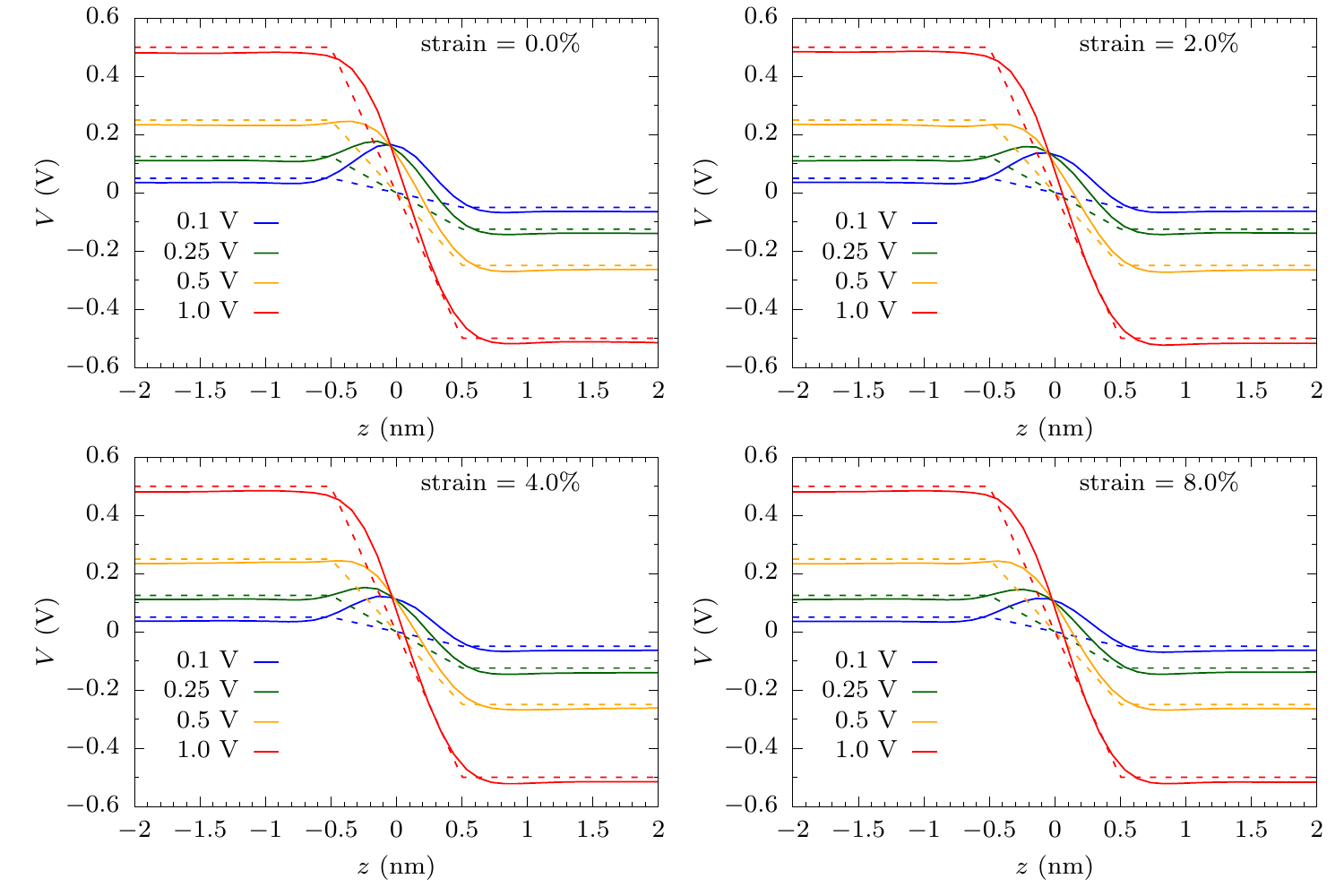}
    \caption{Potential drop along the $z-$axis in the pores with $\qO=-0.24\,e$ and various strains for different applied voltages. The dashed line shows the model where $V$ drops uniformly between $|z|\leq 0.5$ nm and constant outside it. 
    \label{fig:V_q0.24}}
\end{figure}

\begin{figure}[h]
    \centering
    \includegraphics{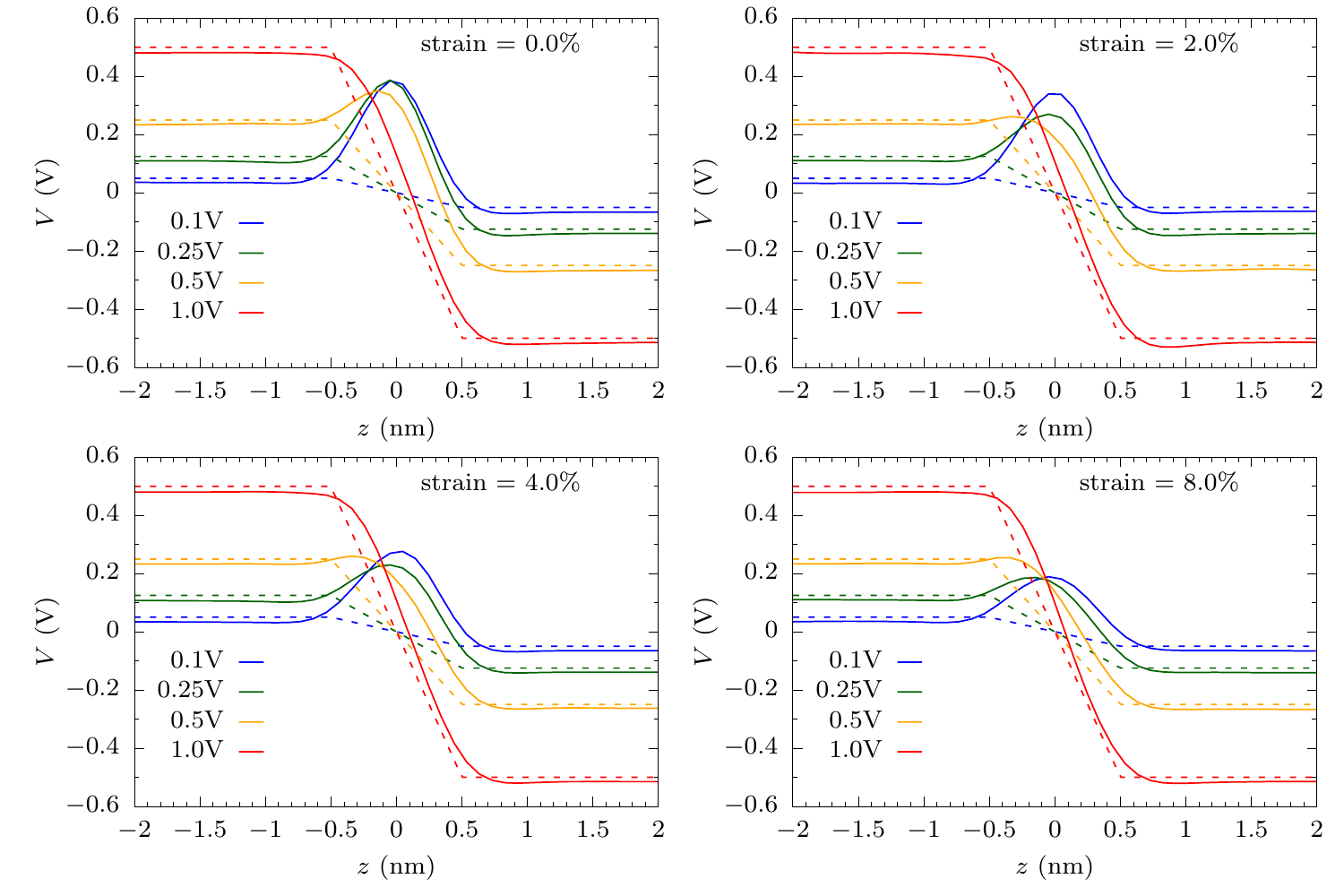}
    \caption{Potential drop along the $z-$axis in the pores with $\qO=-0.54\,e$ and various strains for different applied voltages. The dashed line shows the model where $V$ drops uniformly between $|z|\leq 0.5$ nm and constant outside it. 
    \label{fig:V_q0.54}}
\end{figure}

\begin{figure}[h]
\centering
\includegraphics[width=\textwidth]{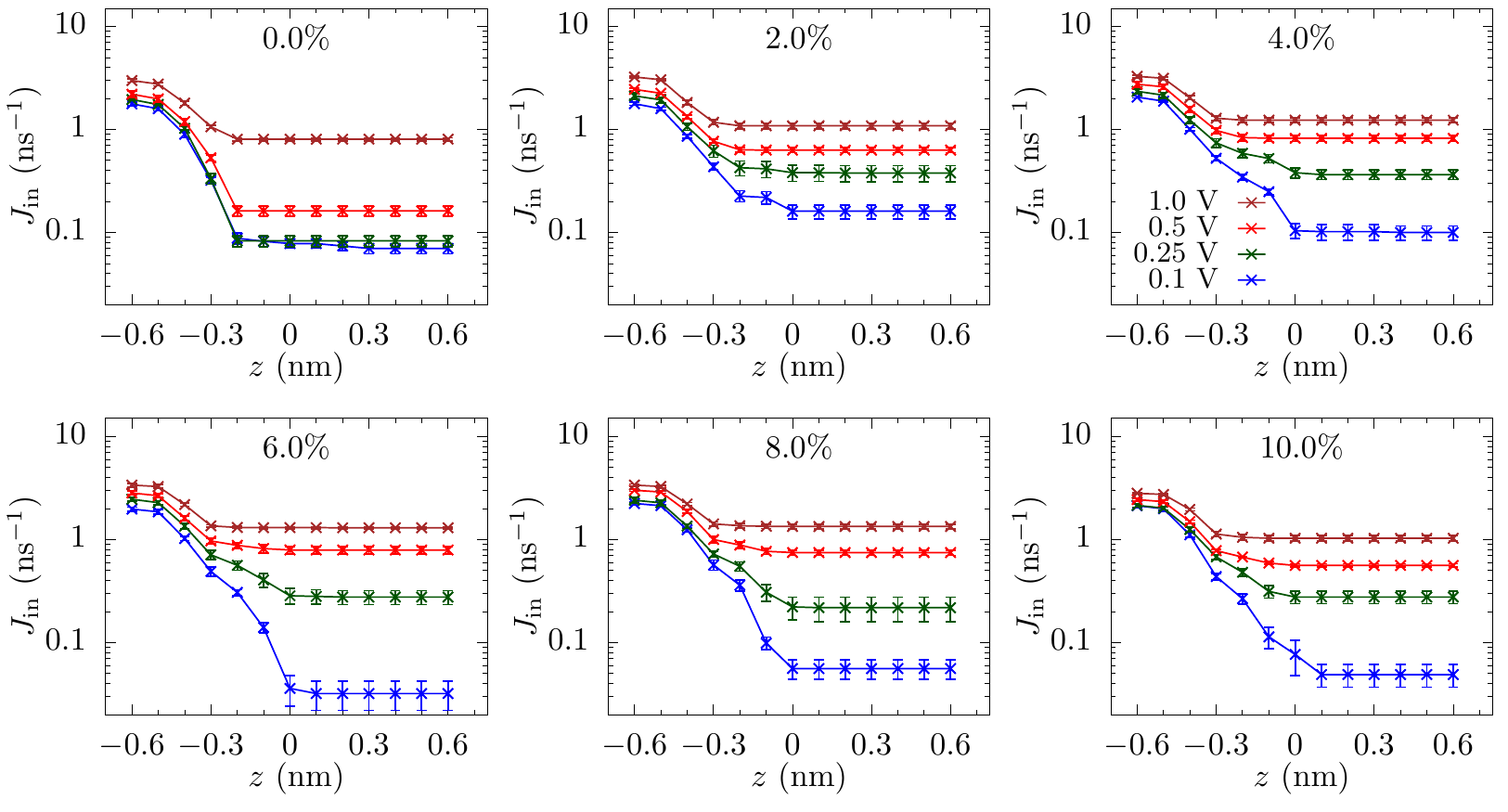}
\caption{The inward flux of K$^+$ ions versus $z$-distance at different applied voltage for pore with $\qO=-0.24\,e$. The error bars are plus/minus one SE from five parallel runs. \label{fig:kinq0.24}}
\end{figure}

\begin{figure}[h]
\centering
\includegraphics[width=\textwidth]{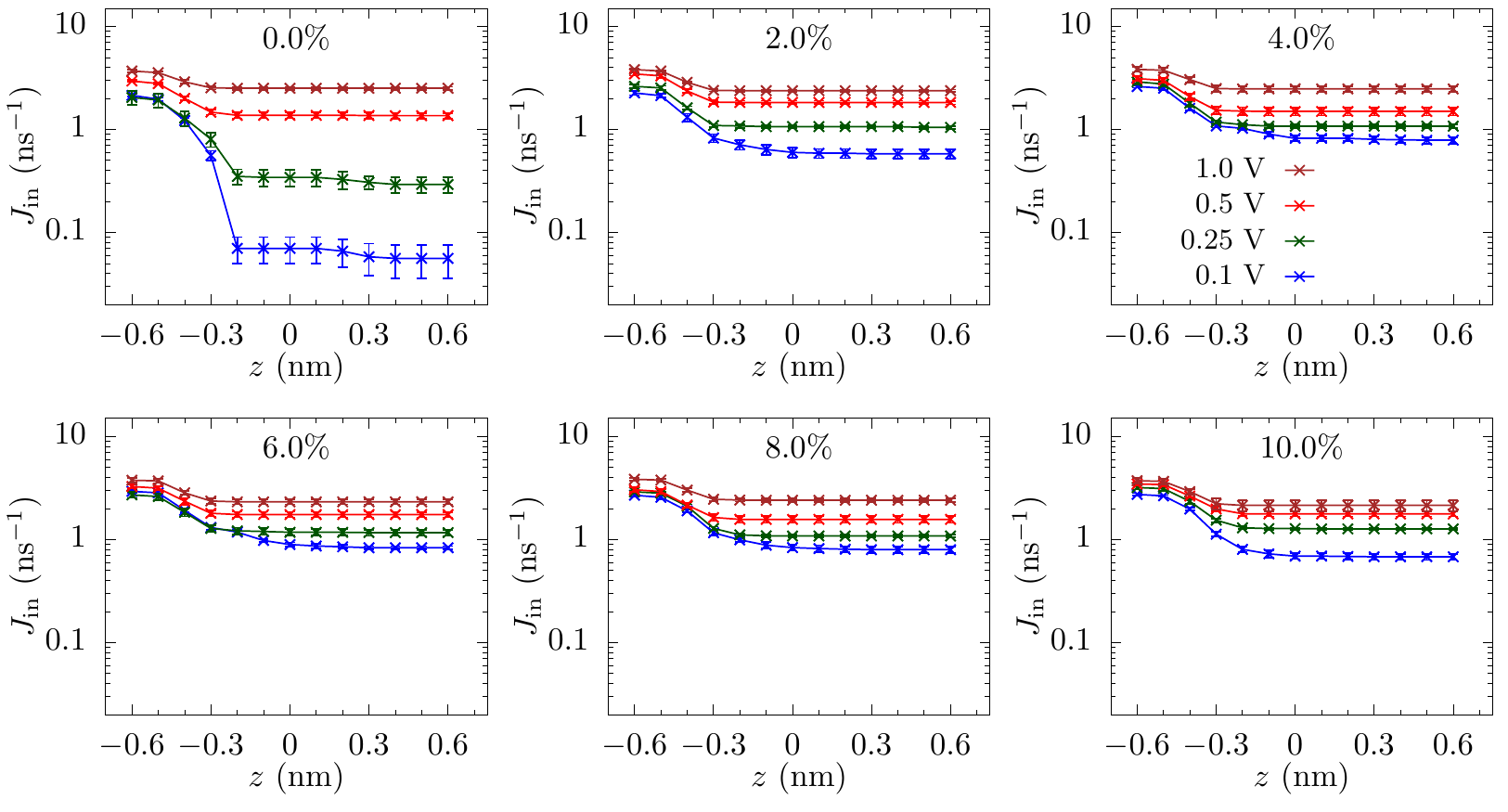}
\caption{The inward flux of K$^+$ ions versus $z$-distance at different applied voltage for pore with $\qO=-0.54\,e$. The error bars are plus/minus one SE from five parallel runs. \label{fig:kinq0.54}}
\end{figure}

\begin{figure}[h]
\centering
\includegraphics[width=\textwidth]{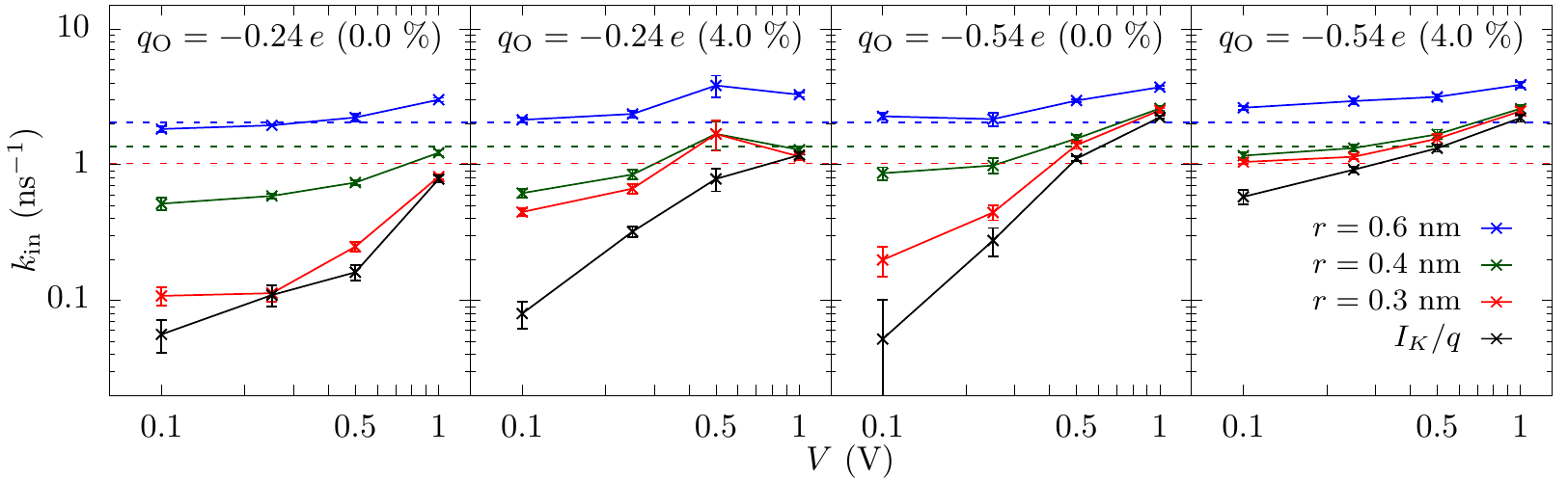}
\caption{Diffusion and barrier limited currents.
The inward diffusion rate of K$^+$ at a distance 0.6~nm, 0.4~nm, and 0.3~nm from the pore versus applied voltage. The dashed horizontal lines give the rate from the diffusion equation assuming ions only enter through a quarter of the sphere (i.e., $\pi\,D\,c\, r$). For small voltage, the $k_\text{in}$ near the pore is much smaller than diffusion limit and thus the current is limited by the barrier to transport. At large voltage, $k_\text{in}$, and hence the current, approach the diffusion limit.  The error bars are plus/minus one SE from five parallel runs.
\label{fig:kin-V}}
\end{figure}

\clearpage
\prg{Other: }
In the remaining figures of the SM, we show additional one-way rate data. Figures~\ref{fig:kinq0.24} and~\ref{fig:kinq0.54} show the one-way rate data as in the main text (across $z$-planes) for the rest of the parameter regimes. Figure~\ref{fig:kin-V} instead shows one-way rate data across hemispherical surfaces.

